%% file: JNW16.tex
\documentclass
      [aip,jcp,preprint,amsmath,amssymb,superscriptaddress,groupedaddress]{revtex4-1}

\usepackage{graphicx} 
\usepackage{longtable}
\usepackage{color}
\usepackage{booktabs}
\AtBeginDocument{
	\heavyrulewidth=.08em
	\lightrulewidth=.05em
	\cmidrulewidth=.03em
	\belowrulesep=.65ex
	\belowbottomsep=0pt
	\aboverulesep=.4ex
	\abovetopsep=0pt
	\cmidrulesep=\doublerulesep
	\cmidrulekern=.5em
	\defaultaddspace=.5em
}
\usepackage{siunitx}
\usepackage{etoolbox}
\robustify\bfseries
\robustify\itshape

\input{mydefs}

\begin{document}

\title{Non-additive Non-interacting Kinetic Energy of Rare Gas Dimers}

\author{Kaili Jiang}
\affiliation{Department of Physics and Astronomy, Purdue University, 525 Northwestern Ave., West Lafayette, IN 47907, USA}

\author{Jonathan Nafziger}
\affiliation{Department of Chemistry, Purdue University, 560 Oval Dr., West Lafayette IN 47907, USA}

\author{Adam Wasserman}
\email[Corresponding Author: ]{awasser@purdue.edu}
\affiliation{Department of Chemistry, Purdue University, 560 Oval Dr., West Lafayette IN 47907, USA}
\affiliation{Department of Physics and Astronomy, Purdue University, 525 Northwestern Ave., West Lafayette, IN 47907, USA}


%
\begin{abstract}
Approximations of the non-additive non-interacting kinetic energy (NAKE) as an explicit functional of the density are the basis of several electronic structure methods that provide improved computational efficiency over standard Kohn-Sham calculations.  
However, within most fragment-based formalisms, there is no unique {\em exact} NAKE, making it difficult to develop general, robust approximations for it. 
When adjustments are made to the embedding formalisms to guarantee uniqueness, approximate functionals may be more meaningfully compared to the exact unique NAKE.  
We use numerically accurate inversions to study the exact NAKE of several rare-gas dimers within Partition Density Functional Theory, a method that provides the uniqueness for the exact NAKE.  
We find that the NAKE decreases nearly exponentially with atomic separation for the rare gas dimers.  
We compute the logarithmic derivative of the NAKE with respect to the bond length for our numerically accurate inversions as well as for several approximate NAKE functionals.  
We show that standard approximate NAKE functionals do not reproduce the correct behavior for this logarithmic derivative, and propose two new NAKE functionals that do. The first of these is based on a re-parametrization of a conjoint PBE functional. The second is a simple, physically-motivated non-decomposable NAKE functional that matches the asymptotic decay constant without fitting.

\end{abstract}

\maketitle

\section{Introduction}

Since their initial development in the 90's, density embedding methods \cite{Cor91,WW93,JN14} have spurred interest in kinetic energy (KE) density functionals. 
In these methods, a system is partitioned into fragments, and an embedding potential is introduced to compensate for the fragment-fragment interactions. 
The total energy of the system is recovered from the sum of fragment energies plus an interaction term.  
The embedding potential for a given fragment is related to the functional derivative of this interaction energy.  
For embedding calculations that use standard density functionals, the Hartree, exchange, correlation and electron-nuclear terms of the interaction energy can be written exactly as functionals of the density.  
However, the KE component must be approximated if it is to be written as an explicit density functional.  
This component is known as the non-additive non-interacting kinetic energy (NAKE) and it is a critical component in describing these fragment-fragment interactions.  
It is defined as the difference between the non-interacting KE of the entire system density and the sum of the non-interacting KE of the individual fragment densities:
\begin{equation}\label{eqn:Tsnad}
	T\s\nad[\{n_\alpha\}]=T\s[n]-\sum_{\alpha}T\s[n_\alpha]
\end{equation}
where $T\s\nad[\{n_\alpha\}]$ is the NAKE written as a functional of the set of fragment densities $\{n_\alpha(\br)\}$, which sum to the total density $n(\br)$.
With any approximation for $T\s[n]$, one can get an approximation for $\Tsnad$ via Eq. (\ref{eqn:Tsnad}).
The approximations obtained in this way are called {\em decomposable} approximations \cite{JN14}.

One general approach to evaluate and improve density functional methods is to numerically evaluate the NAKE as an implicit density functional.  
A few methods have already been developed to accurately evaluate the exact $\Tsnad$ as well as its functional derivative.
These methods typically involve computationally expensive inversion methods\cite{FJN+10,GAMMI10,HPC11,NWW11,NJW17,JW17}.

The downside of using these highly accurate NAKE methods to evaluate the performance of approximate density functionals within standard subsystem DFT is that the solutions are no longer uniquely determined by a given partitioning of a molecule. 
In other words, for a given partitioning, there are multiple possible values of the NAKE that will reproduce the same total energy of the system. 
This non-uniqueness is removed in Partition Density Functional Theory (P-DFT) \cite{CW07,EBCW10,NW14},
where the partitioning is done by minimizing the sum of the fragment energies under the constraint that the sum of the fragment densities matches the total molecular density. 
The uniqueness can also be restored within subsystem-DFT by imposing the additional constraint that all fragments share the same embedding potential \cite{HPC11}.

In the first part of this work, we look at the behavior of approximate NAKE functionals alongside highly accurate numerical evaluations of the NAKE for the rare-gas dimers.  
We use P-DFT to ensure uniqueness of the resulting fragments.  We then compare the behavior of our highly accurate implicit functional with the behavior of approximate functionals.
In the second part, we will introduce a new functional (\Repar), a conjoint PBE functional re-parameterized with a set of NAKE data, and demonstrate that it improves the ground-state energy and density as well as the binding curve in P-DFT calculations for the rare-gas dimers. 
In the third part, we will introduce a physically motivated, {\em non-decomposable} NAKE functional that accurately reproduces the asymptotic behavior for the rare-gas dimers.

\section{Partition-DFT}

A detailed review of P-DFT can be found in Ref. \cite{NW14}.
Here is a brief summary.
In P-DFT, the system is partitioned into non-interacting fragments by dividing the total potential $v(\br)$ into fragment potentials $v_\alpha(\br)$, where $\alpha$ is the fragment index. 
Each fragment is assigned $N_\alpha$ electrons.

The sum of the fragment energies does not match the energy of the whole system.
The remaining part is defined as the partition energy
\begin{equation}\label{eqn:Ep}
	E\p[n]=E[n]-\sum_{\alpha}E_\alpha[n_\alpha]
\end{equation}
Minimization of the sum of the fragment energies under the constraint that the sum of the fragment densities equals the molecular density leads to the fragment KS equations:
\begin{equation}
	\left\{-\frac{1}{2}\nabla^2+v_{\sss{s},\alpha}[n_{\alpha}](\br)+v\p[\{n_{\alpha}\}](\br)\right\}\phi_{i,\alpha}(\br)=\epsilon_{i,\alpha}\phi_{i,\alpha}(\br)
\end{equation}
where the partition potential $v\p[\{n_{\alpha}\}](\br)$ is the Lagrange multiplier associated with the constraint. 
It can be shown that in order to satisfy the density constraint, the partition potential must be set equal to the functional derivative of the partition energy with respect to $n(\br)$:
\begin{equation}\label{eqn:vp}
	v\p(\br)=\dfrac{\delta E\p[n]}{\delta n(\br)}
\end{equation}
The partition energy may be broken into components,
\begin{equation}
\begin{aligned}
\label{eqn:Ep_comp}
	E\p[\{n_{\alpha}\}] = T\s^{\rm nad}[\{n_{\alpha}\}]+&V\ext^{\rm nad}[\{n_{\alpha}\}]\\
	+ E\H^{\rm nad} [\{n_{\alpha}\}] +&E\xc^{\rm nad}[\{n_{\alpha}\}]~~,\\
\end{aligned}
\end{equation}
where each component can be expressed exactly as an explicit functional of the fragment densities, except for the KE component.  
In previous work\cite{NJW17} we developed an inversion procedure capable of numerically evaluating $\Tsnad$.  
With this method, we can reproduce the KS-DFT energies to within $10^{-8}$ Hartree.  
We refer readers to this paper for more details on the inversion procedure.

\section{Approximate functionals for $T\s[n]$}

Semilocal approximations for $T\s[n]$ can be written in the general form
\begin{equation}\label{eqn:Fs}
	T\s[n]=\frac{3}{10}(3\pi^2)^{2/3}\intdhr n(\br)^{5/3}F(s)
\end{equation}
where $s=|\nabla n(\br)|/(k{\sss{F}}n(\br))$ is the reduced density gradient with $k{\sss{F}}=(3\pi^2n(\br))^{1/3}$ and $F(s)$ is an enhancement factor.
There are many strategies to develop approximate KE functionals that can be written in the form provided by Eq. (\ref{eqn:Fs}) with only different forms of $F(s)$. 
For instance, the local density approximation of the KE proposed by Thomas and Fermi (TF) \cite{Tho26,Fer27} has the simplest form of the enhancement factor with $F\TF(s)=1$. 
Another explicit KE functional developed by von Weizs{\"a}cker (vW) \cite{Wei35} has the enhancement factor $F\VW(s)=\frac{5}{3}s^2$.

One strategy to construct KE functionals is through a linear combination of TF and vW.
They can be written in the general form \cite{TW13}
\begin{equation}
\label{eqn:enhancement}
	F(s)=F\TF(s) + \lambda F\VW(s)~~.
\end{equation}
In this paper, we consider the choices of $\lambda$ listed in Table \ref{tab:tfvw} \cite{KP56,Kir57a,Gol57,YT65,Bal72,Lie81}, and refer to this type of KE functionals as \mTFVW functionals.

\begin{table*}[htp]
	\begin{tabular}{l c}
		\toprule
		{Functional} & {$\lambda$}\\
		\midrule
		\TFVW	& 1	\\
		\GEt	& 1/9	\\
		\GOLDEN	& 13/45	\\
		\YTsf	& 1/5	\\
		\BALTIN	& 5/9	\\
		\LIEB	& 0.18590919	\\
		\bottomrule

	\end{tabular}
	\caption{Choice of $\lambda$ of \mTFVW functionals defined by the enhancement factor of Eq. (\ref{eqn:enhancement})}
	\label{tab:tfvw}
\end{table*}

Another strategy to construct KE functionals is based on the \textquotedblleft{conjointness conjecture}\textquotedblright\cite{MS91,LLP91}.
Many KE functionals conjoint the PBE \cite{PBE96} exchange energy functional, which has the enhancement factor 
\begin{equation}\label{eqn:pbe}
	F\PBE(s)=1+\kappa-\frac{\kappa}{1+\frac{\mu}{\kappa}s^2}
\end{equation}
The choice of parameters $\kappa$ and $\mu$ of the conjoint PBE functionals considered in this paper are listed in Table \ref{tab:pbe} \cite{TW02a,CFLDS11,LFCDS11}. 
Other KE functionals that use the conjointness conjecture strategy include \LLP \cite{LLP91}, \FRBee \cite{FR95}, \FRPWes \cite{FR95}, \LCnr \cite{LC94} and \THAKKAR \cite{Tha92}.

\begin{table*}[htp]
	\begin{tabular}{ l S[table-format=1.4] S[table-format=1.5]}
		\toprule
		{Functional} & {$\kappa$} & {$\mu$}\\ 
		\midrule
		\TWa	& 0.8209	& 0.2335	\\
		\TWb	& 0.6774	& 0.2371	\\
		\TWc	& 0.8438	& 0.2319	\\
		\TWd	& 0.8589	& 0.2309	\\
		\APBE	& 0.8040	& 0.23889	\\
		\REVAPBE	& 1.245 & 0.23889	\\
		\APBEINT	& 0.8040	& {$\dfrac{5/9+5s^2\cdot0.23889}{3+5s^2}$}	\\
		\REVAPBEINT	& 1.245	& {$\dfrac{5/9+5s^2\cdot0.23889}{3+5s^2}$}	\\
		\bottomrule
	\end{tabular}
	\caption{Choice of $\kappa$ and $\mu$ of conjoint PBE functionals.}
	\label{tab:pbe}
\end{table*}

There are of course other choice of approximate functionals for $T\s[n]$ not considered in this work \cite{WGC99,HC10,XC15,KT15,CFDS17,GP17,TW13}.

\section{Behavior of numerically exact $T\s\nad$ and its derivatives}
We calculated highly accurate NAKEs at a wide range of separations for the rare-gas dimers using our own all-electron real-space code CADMium.
LDA is used as the exchange-correlation functional throughout this paper, and the equilibrium stands for the LDA equilibrium separation. The LDA is known to severely overestimate the binding energy of noble-gas dimers \cite{KH15}. This is due almost entirely to functional-driven errors, not density-driven errors \cite{WNJ+17}, so the conclusions of our work will only depend minimally on this choice of XC functional. The exact NAKE in this paper is the one that reproduces the LDA results, including all of its errors.
Table \ref{tab:rare-gas} provides a comparison for each rare-gas dimer between P-DFT and KS-DFT calculations performed on the same grid using CADMium and with KS results from NWChem \cite{VBG+10}.
This table shows that the differences between the P-DFT and the KS-DFT binding energies from CADMium are in the order of $10^{-8}$ Hartree or less in all cases.
The differences between the KS-DFT binding energies from CADMium and from NWChem are in the order of $10^{-5}\sim10^{-6}$ Hartree, which is mostly due to the difference between using finite basis sets (NWChem) and a real-space grid (CADMium). 
We use the NAKE generated with the inversion procedures in CADMium as benchmark for the numerically \textquotedblleft exact\textquotedblright\xspace LDA NAKE in this paper.

\begin{table*}[htp]
	\begin{tabular}{l *{5}{S[table-format=2.4]}}
		\toprule
		{System}     & 
		{\begin{tabular}[c]{@{}c@{}}$E_{\sss{b,KS}}$\\ (mHa)\end{tabular}} & {\begin{tabular}[c]{@{}c@{}}$E_{\sss{b,inv}}$\\   (mHa)\end{tabular}} & 
		{\begin{tabular}[c]{@{}c@{}}$E_{\sss{b,KS}}-E_{\sss{b,inv}}$\\ (nHa)\end{tabular}} &
		{\begin{tabular}[c]{@{}c@{}}$E_{\sss{b,NWChem}}$\\   (mHa)\end{tabular}} &
		{\begin{tabular}[c]{@{}c@{}} $E_{\sss{b,KS}}-E_{\sss{b,NWChem}}$\\ ($\mu$Ha)\end{tabular}} \\
		\midrule
		He$_2$	   & -0.3550 & -0.3550 & -13.5040 & -0.3591 & 4.0604   \\
		Ne$_2$ 	   & -0.7372 & -0.7372 & 11.6920  & -0.7847 & 47.5650  \\
		Ar$_2$ 	   & -1.1175 & -1.1175 & -15.4400 & -1.1186 & 1.1200   \\
		HeNe   	   & -0.5495 & -0.5494 & -9.1450  & -0.5653 & 15.8160  \\
		HeAr   	   & -0.5442 & -0.5442 & 3.2844   & -0.5489 & 4.7381   \\
		NeAr   	   & -0.8770 & -0.8770 & 11.2960  & -0.8994 & 22.3530  \\
		\bottomrule	
	\end{tabular}
	\caption{Comparison of binding energy at the equilibrium using KS-DFT and P-DFT with inversion calculations in CADMium, and KS-DFT calculations from NWChem. The aug-cc-pvtz basis set is used for NWChem calculations.}
	\label{tab:rare-gas}
\end{table*}

\subsection{Asymptotic Behavior} \label{sec:asympt_exact}

\begin{figure}[htb]
	\includegraphics*[width=3.375in]{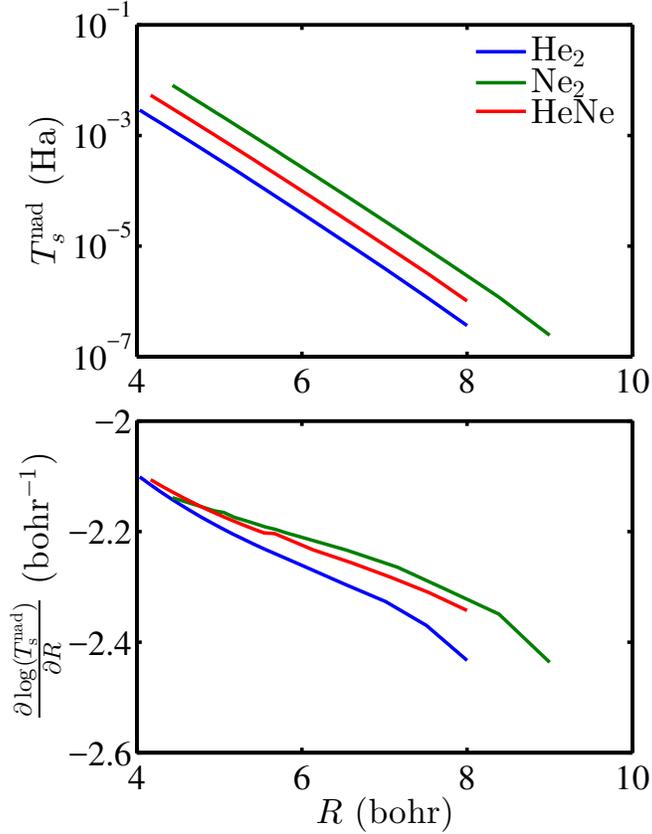}
	\caption{Top: The LDA NAKE {\em vs.} inter-nuclear separation for three rare-gas dimers. Bottom: logarithmic derivative of the NAKE {\em vs.} inter-nuclear separation for the same systems.}
	\label{fig:rare-gas}
\end{figure}

Figure \ref{fig:rare-gas} panel 1 shows the LDA NAKE {\em vs.} inter-nuclear separation ($R$) for He$_2$, Ne$_2$ and HeNe. 
It is apparent that for each of these systems the NAKE is always positive and behaves as a nearly exponential function of the separation.  
In order to explore this behavior further, we numerically calculate the logarithmic derivative of the NAKE {\em vs.} the separation (Figure \ref{fig:rare-gas} panel 2). 
A perfectly exponential function would have a constant logarithmic derivative, but the figure shows the logarithmic derivative varying in a small range:
The NAKE is a nearly exponential function of $R$, i.e. $\Tsnad\sim e^{-kR}$. The best fit to the exponential decay constant ($k$) is reported in Table \ref{tab:decay_constant}.

\begin{table*}[htp]
	\begin{tabular}{l S[table-format=1.4]}
		\toprule
		{System} & {Decay constant}   \\
		\midrule
		He$_2$ & 2.2508 \\
		Ne$_2$ & 2.2549 \\
		Ar$_2$ & 1.8701 \\
		HeNe   & 2.2277 \\
		HeAr   & 1.9681 \\
		NeAr   & 1.9837 \\
		\bottomrule
	\end{tabular}
	\caption{The decay constant $k$ of the rare-gas dimers obtained by fitting the NAKE data into $\Tsnad=Ce^{-kR}$. Units in $\mathrm{bohr}^{-1}$.}
	\label{tab:decay_constant}
\end{table*}

\subsection{Unambiguous NAKE per particle}

The NAKE per particle $\epkin(\br)$ measures the performance of the KE functionals in different regions.
It is defined as
\begin{equation}
	\epkin(\br) = \dfrac{\tau\nad(\br)}{n(\br)}
\end{equation}
where $\tau\nad(\br)$ is a NAKE density satisfying
\begin{equation}\label{eqn:tau_nad}
	\Tsnad = \intdhr \tau\nad(\br)
\end{equation}

Similar to the NAKE, $\tau\nad(\br)$ can be written as the difference between the KE density of the entire system and that of the fragments:
\begin{equation}
	\tau\nad(\br)=\tau(\br)-\sum_{\alpha}\tau_\alpha(\br)
\end{equation}
where $\tau(\br)$ and $\tau_\alpha(\br)$ are the NAKE density for the entire system and for fragment $\alpha$, respectively.

The KE density $\tau(\br)$ is not uniquely defined, as any function that integrates to zero over all space, e.g., $\nabla^2 n(\br)$, can be added to a valid $\tau(\br)$ to produce another equally valid KE density \cite{SLBB03}. The non-uniqueness is partially removed for the NAKE density as long as the same form of KE density is employed for both the fragments and the entire system because the laplacian is a linear operator: $\nabla^2 n(\br)=\sum_{\alpha}\nabla^2n_\alpha(\br)$.

The reason of using the NAKE per particle $\epkin(\br)$ instead of the NAKE density $\tau\nad(\br)$ is that $\tau\nad(\br)$ is very localized in the region of the nuclei, while $\epkin(\br)$ has more delocalized features.


\begin{figure}[htb]
	\includegraphics*[width=\textwidth]{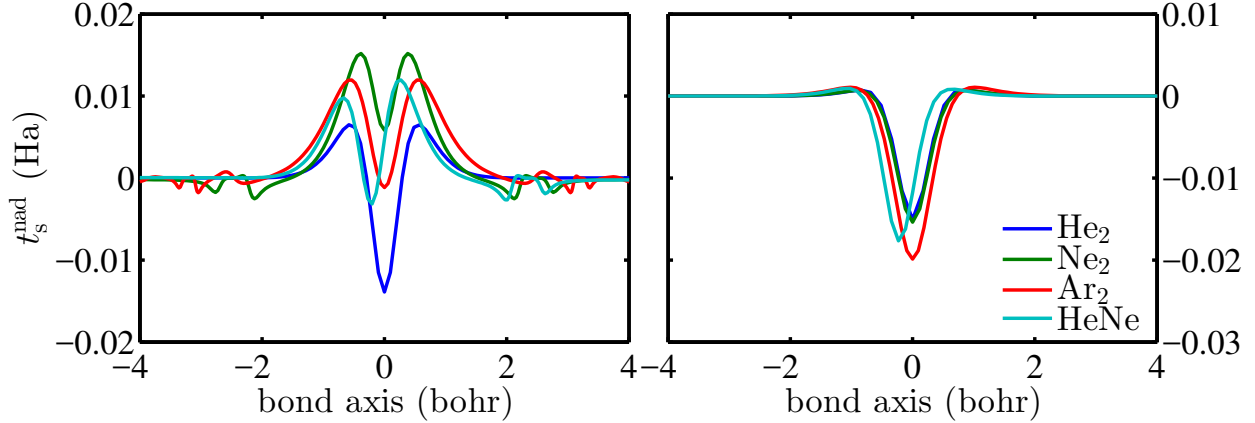}
	\caption{The numerically exact LDA NAKE per particle on the bond axis for select rare-gas dimers. Left: equilibrium. Right: large separation.}
	\label{fig:rare_gas_epkin}
\end{figure}

Figure \ref{fig:rare_gas_epkin} shows that the exact NAKE per particle for the rare-gas dimers has a central feature of a double peak with a well in the middle. 
As the separation of the nuclei of the dimers increases, the peaks move farther apart and become much smaller, while the well becomes wider and deeper.
Also, small features exist in the region of the Ne and Ar nuclei.
Those features are visible at the equilibrium but can be hardly seen at larger separations.

\subsection{Non-additive non-interacting kinetic potential (NAKP)}
Any effective approximation to the NAKE must perform well in two aspects, its value and its functional derivative \cite{BW17}. 
The latter is called NAKP $\vpkin(\br)$:
\begin{equation}\label{eqn:vpkin]}
	\vpkin(\br)=\dfrac{\delta\Tsnad}{\delta n(\br)}
\end{equation}
It has been shown that the improvements in approximations of the NAKE do not necessarily lead to the improvements in approximations of the NAKP \cite{GBV09,WS13}.
Therefore, it is necessary to study the behavior of the NAKP.

In P-DFT, the NAKP is unique up to a constant.
Therefore, P-DFT provides a suitable framework to analyze the behavior of the NAKP for various approximate KE functionals.



\begin{figure}[htb]
	\includegraphics*[width=\textwidth]{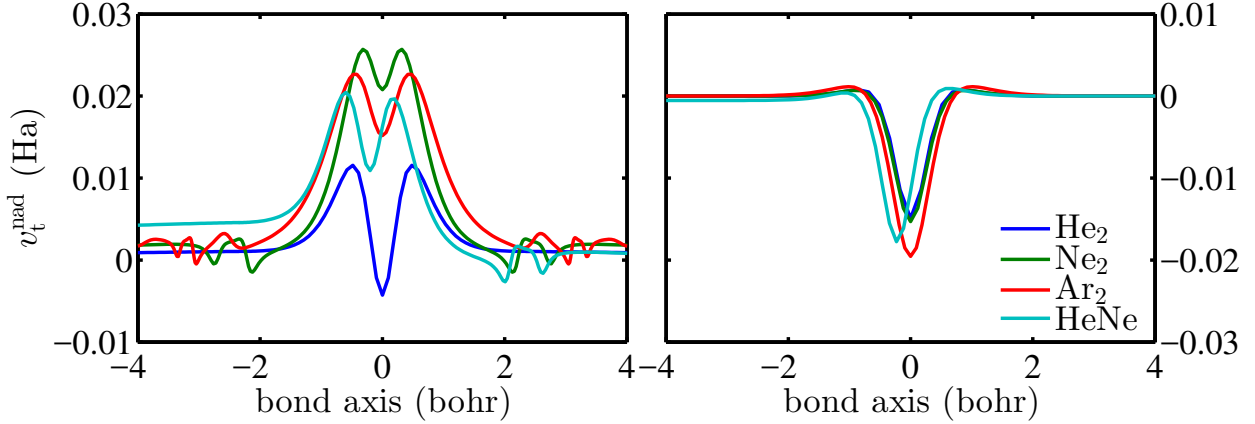}
	\caption{The exact NAKP on the bond axis for rare-gas dimers. Left: equilibrium. Right: large separation.}
	\label{fig:rare_gas_vpkin}
\end{figure}

Similar to the behavior of the NAKE per particle, figure \ref{fig:rare_gas_vpkin} shows that the exact NAKP also has the feature of a double peak with a single well for all the systems we considered. 
The depth of the wells is smaller compared to that of the NAKE per particle at the equilibrium, but of about the same size at the larger separations.
The exact NAKP also shows some small features in the region of the Ne and the Ar nuclei. 
The difference in the level between the left and the right side of asymmetrical systems is due to the difference in the chemical potential between the two atoms.

\section{Re-parametrization}

Most KE functionals fit the parameters to a set of $T\s[n]$ data.
In this work, we choose the PBE form of the enhancement factor in Eq. \ref{eqn:pbe} and re-parameterize it by fitting to our accurate set of $\Tsnad$ data for rare-gas dimers.

The re-parametrization is done in the following way. 
For each rare-gas dimer, we choose 13 different separations evenly distributed between ${R\sss{eq}}-1/4\mathrm{bohr}$ and ${R\sss{eq}}+1/2\mathrm{bohr}$, where ${R\sss{eq}}$ is the equilibrium separation. 
We then look for the $\kappa$ and $\mu$ that minimize the error $\Delta$
\begin{equation}\label{eqn:min_repar}
	\Delta = \sum_{\mathrm{Systems}}\sum_{R_i}\left(\frac{\tilde{T}_{\sss{s},R_i}\nad-T_{\sss{s},R_i}\nad}{T_{\sss{s},R_i}\nad}\right)^2
\end{equation}
where $\tilde{T}_{\sss{s},r_i}\nad$ is the non-self-consistent NAKE evaluated from the exact P-DFT density using the re-parameterized enhancement factor and $R_i$ is the $i$th separation.

\begin{table*}[htp]
	\begin{tabular}{l S[table-format=1.4] S[table-format=1.5]}
		\toprule
		{Fitting set} & {$\kappa$} & {$\mu$}\\ 
		\midrule
		He$_2$              & 2.0654 & 0.01042 	\\
		Ne$_2$              & 2.3234 & 0.02748 	\\
		Ar$_2$              & 2.2049 & 0.01906 	\\
		HeNe             	& 2.0158 & 0.02052 	\\
		HeAr              	& 2.0301 & 0.01303 	\\
		NeAr             	& 2.0777 & 0.02808 	\\
		He$_2$, Ne$_2$, Ar$_2$, HeNe, HeAr, NeAr 	& 1.9632 & 0.01979	\\
		\bottomrule
	\end{tabular}
	\caption{Parameters $\kappa$ and $\mu$ obtained by fitting into the NAKE data of rare-gas dimers}
	\label{tab:repar}
\end{table*}

The results are shown in Table \ref{tab:repar}. 
The optimal value of $\kappa$ is close to 2 for all systems, which is much larger than the typical $\kappa$ value of the conjoint PBE functionals listed in table \ref{tab:pbe}. 
Laricchia \textit{et al.}\cite{LFCDS11} suggest that a large value of $\kappa$ in \REVAPBE is needed to obtain improved embedding energies in frozen density embedding calculations. 
Our results show that to obtain accurate NAKE in P-DFT calculations, the value of $\kappa$ needs to be even larger than that of \REVAPBE. 
On the other hand, the optimal value of $\mu$ is much smaller and varies between $0.01-0.03$ for different systems. 
The last set of $\kappa$ and $\mu$ in Table \ref{tab:repar} is named \Repar.

\subsection{Scaling separation distance}


\begin{figure}[htb]
	\includegraphics*[width=\textwidth]{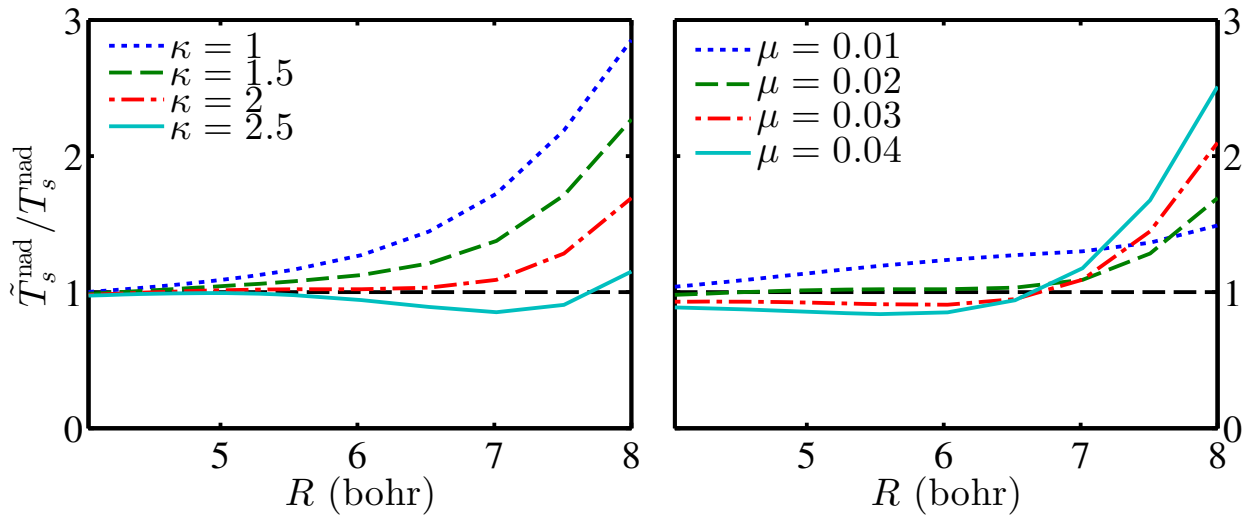}
	\caption{The ratio of the approximated and the exact NAKE of different choices of $\kappa$ and $\mu$ in the PBE enhancement factor versus the inter-nuclear separation for He$_2$. Left: fixed $\mu=0.02$. Right: fixed $\kappa=2.0$.}
	\label{fig:binding_rpbe_dis}
\end{figure}

Figure \ref{fig:binding_rpbe_dis} shows how the parameters $\kappa$ and $\mu$ in the PBE enhancement factor affect the asymptotic behavior of the NAKE for He$_2$. 
Overall with a generally wide range of $\kappa$ and $\mu$, the PBE enhancement factor can reproduce the nearly exponential asymptotic behavior except at very large separations. The parameter 
$\kappa$ mostly controls the decay constant. 
As we can see from the figure, the NAKE with different $\kappa$ starts at the same point at small separations, but diverges as the dimer stretches. 
As mentioned in the previous section, in order to reproduce the exact decay constant, a large value of $\kappa$ needs to be chosen. 
On the other hand, $\mu$ shifts the NAKE value up and down near the equilibrium and controls how the curve bends at large separations.
A small value of $\mu$ is required to keep the asymptotic behavior exponential.

\subsection{Behavior of NAKE per particle}
\begin{figure}[htb]
	\includegraphics*[width=\textwidth]{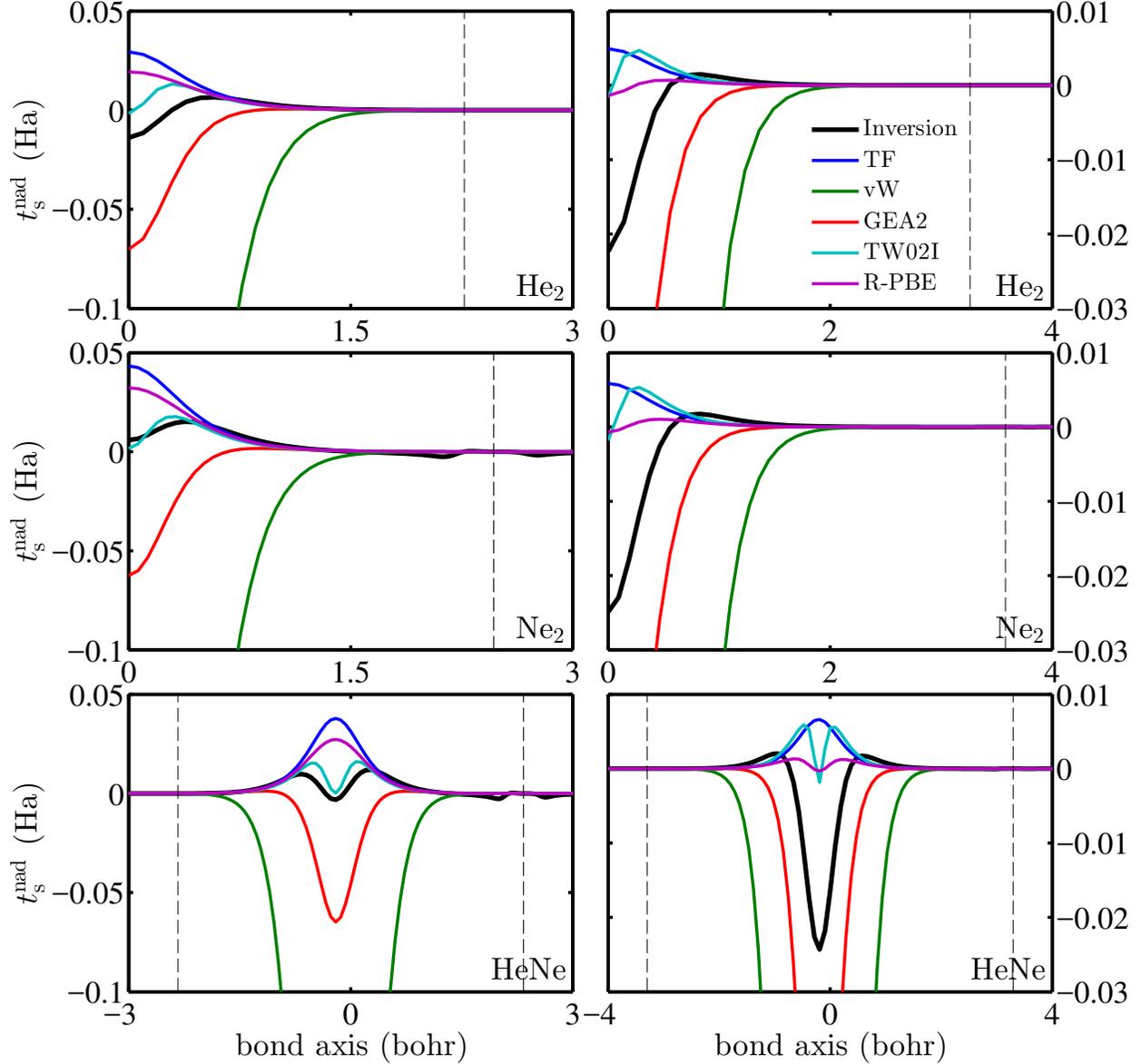}
	\caption{The LDA NAKE per particle for various rare-gas dimers on the bond axis. The left column shows the NAKE per particle at equilibrium and the right column at a larger separation. Only the right half is shown for all systems except HeNe due to symmetry. Dashed lines indicate the locations of nuclei.}
	\label{fig:epkin}
\end{figure}

Figure \ref{fig:epkin} shows the NAKE per particle with selected approximate KE functionals of He$_2$, Ne$_2$ and HeNe on the bond axis.
In general, most of the approximations fail to accurately reproduce the feature of a double peak with a single well, and none of the approximations reproduce the correct features observed around the Ne nuclei. 
\TFa has a single peak and \VWa has a very deep and wide well in the center for both equilibrium and large separations.
However, the \mTFVW functionals do not reproduce the double-peak feature because the well in \VWa is wider than the peak in \TFa.
\TWa does reproduce the double-peak feature, resembling the exact result at equilibrium, but at large separations the peak becomes higher and the well becomes narrower than it should.
Although \Repar fails to reproduce this feature at the equilibrium because the two peaks overlap, it does reproduce the feature at large separations with the well of a more accurate width than \TWa.

\subsection{Behavior of NAKP}
\begin{figure}[htb]
	\includegraphics*[width=\textwidth]{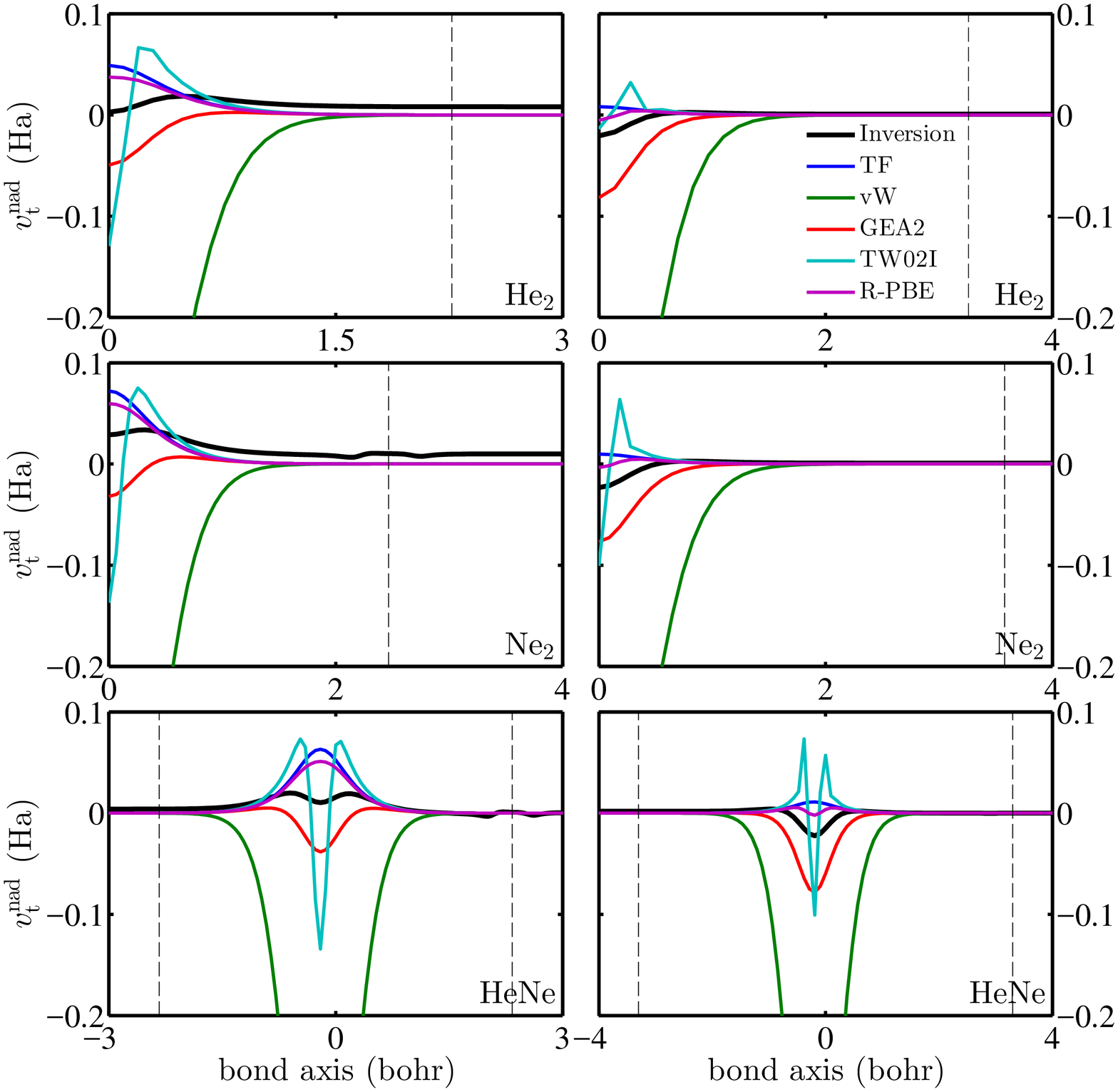}
	\caption{The NAKP for various rare-gas dimers on the bond axis. The left column shows the NAKP at the equilibrium and the right column shows that at a larger separation. Only the right half is shown for all systems except HeNe due to symmetry. Dashed lines indicate the locations of nuclei.}
	\label{fig:vpkin}
\end{figure}

Figure \ref{fig:vpkin} shows the NAKP with selected approximate KE  functionals of He$_2$, Ne$_2$ and HeNe on the bond axis. 
The behavior of the NAKP with approximate functionals is very similar to the behavior of the NAKE per particle of those functionals.
The most noticeable difference is in \TWa.
While it matches the exact result fairly accurately for the behavior of the NAKE per particle, in the case of the NAKP it yields much higher peaks and a much deeper well compared to the inversion results.
On the other hand, \Repar reproduces the feature of a double peak with a single well at large separations, but the width and the depth of the well are smaller than they should be. 
Overall, similar to the behavior of the NAKE per particle, none of the approximate functionals capture the correct behavior at all separations.

\subsection{NAKE and self-consistent density at equilibrium}

In this section, the performance of the KE functionals at the equilibrium is studied for six rare-gas dimers: He$_2$, Ne$_2$, Ar$_2$, HeNe, HeAr and NeAr.
To measure how well an approximate KE functional performs in different regions with only one number, we calculated the error in the self-consistent density using
\begin{equation}
\label{eqn:density_error}
N_{\sss{err}}=\intdhr|n(\br)-n_{\sss{approx}}(\br)|
\end{equation}

Table \ref{tab:err_tsnad} and Table \ref{tab:err_den} show the error in the NAKE and the error in the self-consistent density of selected approximate KE functionals, respectively.
Tables with a full list of approximate functionals can be found in the supplemental data.
The first functional in both tables, which is setting \none, is used as a reference.
A functional should at least perform better than setting \none to be considered as \textquotedblleft{good}\textquotedblright.

\begin{table*}[htp]
	\begin{tabular}{ l *{6}{S[detect-all,table-format=1.1, table-figures-exponent = 1, table-auto-round, scientific-notation=true]}}\\
		\toprule
		{Functional} & {He$_2$} & {Ne$_2$} & {Ar$_2$} & {HeNe} & {HeAr} & {NeAr}\\
		\midrule
		\none       & -0.000992935 & -0.002749803 & -0.004050443 & -0.001844909 & -0.001802383 & -0.003151847 \\
		\TFa        & 0.000204941  & 0.001527064  & 0.000724964  & 0.001057245  & 0.000368284  & 0.001054373  \\
		\VWa        & \itshape -0.038816314 & \itshape -0.078574121 & \itshape -0.081391498 & \itshape -0.058621124 & \itshape -0.053375019 & \itshape -0.079174517 \\
		\GEt        & -0.002553623 & -0.003884723 & -0.004806448 & -0.003068981 & -0.003321866 & -0.004245725 \\
		\Repar      & \bfseries 0.000002206  & 0.000626711  & 0.000260030  & 0.000255579  & \bfseries 0.000081079  & 0.000625061  \\
		\REVAPBE    & -0.000024458 & 0.000714198  & -0.000562756 & 0.000542892  & -0.000128850 & 0.000142425  \\
		\APBE       & 0.000155096  & 0.001209034  & 0.000056365  & 0.000881213  & 0.000180714  & 0.000657349  \\
		\TWa        & 0.000142851  & 0.001177547  & 0.000020797  & 0.000859114  & 0.000160836  & 0.000625420  \\
		\TWb        & 0.000200493  & 0.001346924  & 0.000246111  & 0.000972683  & 0.000266217  & 0.000805601  \\
		\TWc        & 0.000132130  & 0.001147898  & \bfseries -0.000016151 & 0.000838838  & 0.000142280  & 0.000594538  \\
		\TWd        & 0.000125063  & 0.001128482  & -0.000040188 & 0.000825533  & 0.000130121  & 0.000574363  \\
		\FRBee      & -0.000205452 & 0.000510921  & -0.000571979 & 0.000342903  & -0.000289167 & \bfseries -0.000001301 \\
		\LCnr       & -0.000427578 & \bfseries 0.000004412  & -0.001124647 & \bfseries -0.000024549 & -0.000617789 & -0.000507048 \\
		\bottomrule
	\end{tabular}
	\caption{The error in the NAKE of various KE functionals. For each system, the best result is in boldface and the worst is in italics.}
	\label{tab:err_tsnad}
\end{table*}

\begin{table*}[htp]
	\begin{tabular}{ l *{6}{S[detect-all, table-format=1.1, table-figures-exponent = 1, table-auto-round, scientific-notation=true]}}\\
		\toprule
		{Functional} & {He$_2$} & {Ne$_2$} & {Ar$_2$} & {HeNe} & {HeAr} & {NeAr}\\
		\midrule
		\none       & 0.00534874 & 0.01063406 & 0.02142618 & 0.00849323 & 0.01125317 & 0.01452504 \\
		\TFa        & 0.00173069 & 0.00456821 & 0.00735248 & 0.00304807 & 0.00388457 & 0.00558493 \\
		\VWa        & \itshape 0.13640329 & \itshape 0.23635308 & \itshape 0.32698531 & \itshape 0.18914528 & \itshape 0.21060381 & \itshape 0.27557036 \\
		\GEt        & 0.01099064 & 0.01639058 & 0.02328452 & 0.01420872 & 0.01723691 & 0.01945147 \\
		\Repar      & 0.00151018 & 0.00298875 & \bfseries 0.00604696 & 0.00226527 & \bfseries 0.00376447 & \bfseries 0.00445767 \\
		\REVAPBE    & 0.00281139 & 0.00540965 & 0.01119282 & 0.00446332 & 0.00706180 & 0.00826670 \\
		\APBE       & 0.00169300 & 0.00285330 & 0.00749173 & 0.00255087 & 0.00492354 & 0.00553045 \\
		\TWa        & 0.00174258 & 0.00293904 & 0.00759664 & 0.00262676 & 0.00500733 & 0.00562789 \\
		\TWb        & \bfseries 0.00145374 & \bfseries 0.00231568 & 0.00640231 & \bfseries 0.00210744 & 0.00439581 & 0.00472786 \\
		\TWc        & 0.00179784 & 0.00305646 & 0.00777804 & 0.00272134 & 0.00511850 & 0.00577125 \\
		\TWd        & 0.00183482 & 0.00313506 & 0.00789786 & 0.00278474 & 0.00519214 & 0.00586431 \\
		\FRBee      & 0.00262848 & 0.00377747 & 0.00806487 & 0.00357004 & 0.00595793 & 0.00624929 \\
		\LCnr       & 0.00375590 & 0.00586362 & 0.01068964 & 0.00519749 & 0.00766296 & 0.00826116 \\
		\bottomrule
	\end{tabular}
	\caption{The error in the self-consistent density as defined in Eq.(\ref{eqn:density_error}) for various KE functionals. For each system, the best result is in boldface and the worst is in italics.}
	\label{tab:err_den}
\end{table*}

Due to the small overlap between the fragment density at equilibrium, setting \none yields relatively small errors, which is $\sim1\mathrm{mHa}$ for the NAKE, and $\sim0.01$ for the self-consistant density.
All functionals using the conjointness conjecture give an error in the NAKE of less than $1\mathrm{mHa}$ and an error in the density of less thab $0.01$.
In terms of the NAKE, \Repar achieves the best result for He$_2$ and \LCnr for Ne$_2$ and HeNe, while \TWb achieves the best result in terms of the density, followed by \Repar.
It is worth noting that although \REVAPBE yields better results than \APBE in terms of the NAKE as in Ref \cite{LFCDS11}, it yields worse self-consistent densities. 
On the other hand, \VWa yields errors that are $20-40$ times larger than setting \none, making all \mTFVW functionals \textquotedblleft{bad}\textquotedblright.

\subsection{Dissociation}

\begin{figure}[htb]
	\includegraphics*[width=\textwidth]{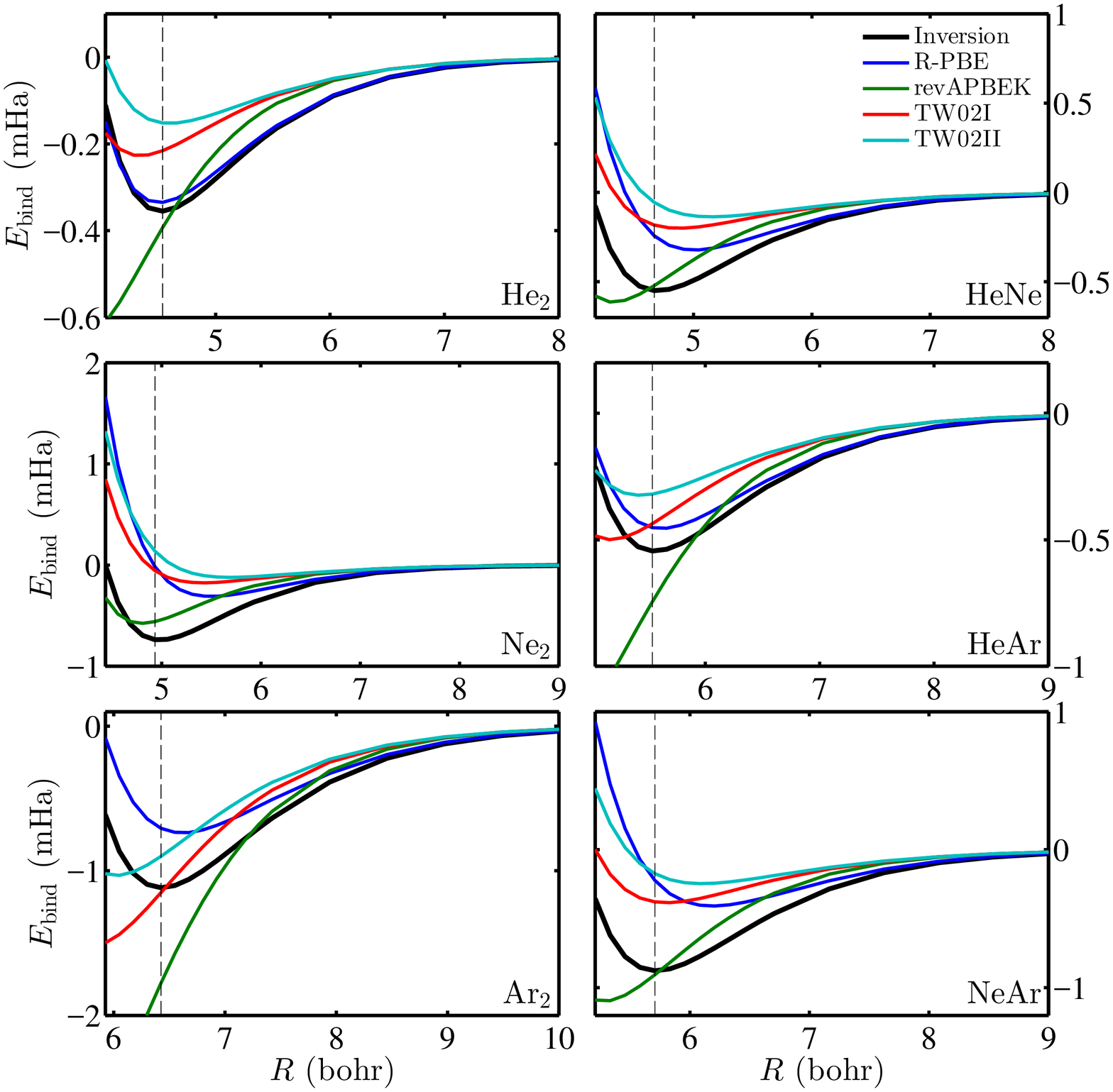}
	\caption{Binding Curves for rare-gas dimers with functionals conjoint PBE used as $\Tsnad$. $R$ is the inter-nuclear separation and $E_\mathrm{bind}$ is the binding energy. Dashed lines indicate the equilibrium separation.}
	\label{fig:binding_pbe}
\end{figure}

The dissociation curve can be used as a measure of the performance of approximate KE functionals as the dimers are stretched.
Overall, conjoint PBE functionals yield better dissociation curves for the rare-gas dimers than other approximate KE functionals.
Figure \ref{fig:binding_pbe} shows the dissociation curve with conjoint PBE functionals for rare-gas dimers.  
\TWa, \TWc, \TWd, \APBE and \APBEINT yield very similar results, so only \TWa is shown, and \REVAPBEINT yields a similar dissociation curve to \REVAPBE.

\Repar errors are smaller than $0.1\mathrm{bohr}$ for the equilibrium bond lengths of He$_2$, Ar$_2$, and HeAr.
For Ne$_2$, HeNe and NeAr, the \Repar equilibrium bond length is larger than the exact result.
\REVAPBE yields an accurate equilibrium bond length for Ne$_2$, \TWa for HeNe and HeAr, and \TWb for He$_2$ and HeAr.
In terms of the binding energy, \Repar matches the exact dissociation curve for He$_2$, and has an error of less than $0.1\mathrm{mHa}$ for HeAr, and it still outperforms the other conjoint PBE functionals for the other rare-gas dimers, especially at large separations.
Although \REVAPBE and \REVAPBEINT have accurate binding energies near the equilibrium, they overbind in all systems, while \TWb underbinds in all systems.
Overall, \Repar performs the best in dissociation for the rare-gas dimers.

\section{Two-orbital approximation (2OA)}

In two-orbital homo-nuclear diatomics one Kohn-Sham orbital has gerade symmetry while the other orbital has ungerade symmetry.  
By treating the fragment densities of these systems as if they represented localized molecular orbitals, we can construct approximations to the gerade and ungerade KS molecular orbitals.  
We begin by studying this idea with non-self-consistent post-P-DFT calculations.  
In this case, each fragment orbital has the same asymptotic behavior as the HOMO, which is the ungerade orbital.  
Because of this, we construct an approximation to the ungerade orbital first:
\begin{equation}\label{eqn:orbital_ug}
\phi_{\rm ug}(\br) \approx N(n_1(\br)^\half-n_2(\br)^\half)
\end{equation}
This approximate orbital will have the correct symmetry and be properly normalized by setting $N=1/(\int(n_1(\br)^\half-n_2(\br)^\half)^2(\br)d^3\br)^\half$.  
After the normalization, we can construct the remaining gerade orbital from the remaining density.
\begin{equation}\label{eqn:orbital_g}
\phi_{\rm g}(\br) \approx \left(\frac{n(\br)}{2}-\phi_{\rm ug}^2(\br)\right)^\half
\end{equation}

This approximation becomes exact as the inter-nuclear separation goes to infinity, but it still does quite well in the bonding region even at relatively short bond lengths.  
These approximate orbitals then lead to approximate KEs that can be used to construct NAKEs:

\begin{equation}\label{eqn:2oa}
T^{\rm nad}\s[n_1,n_2] = -\half\sum_{i={\rm ug,g}} \int{\phi_{i}(\br)\nabla^2\phi_{i}(\br) d\br} + \half\sum_{i={\rm 1,2}} \int{n_{i}^{1/2}(\br)\nabla^2n_{i}^{1/2}(\br) d\br}
\end{equation}

This approximation can only be expected to give reasonable results for He$_2$, but it is interesting to note that it captures the asymptotic behavior of all other rare-gas dimers better than any other approximate functional.

\subsection{Asymptotic Behavior}

\begin{figure}[htb]
	\includegraphics*[width=\textwidth]{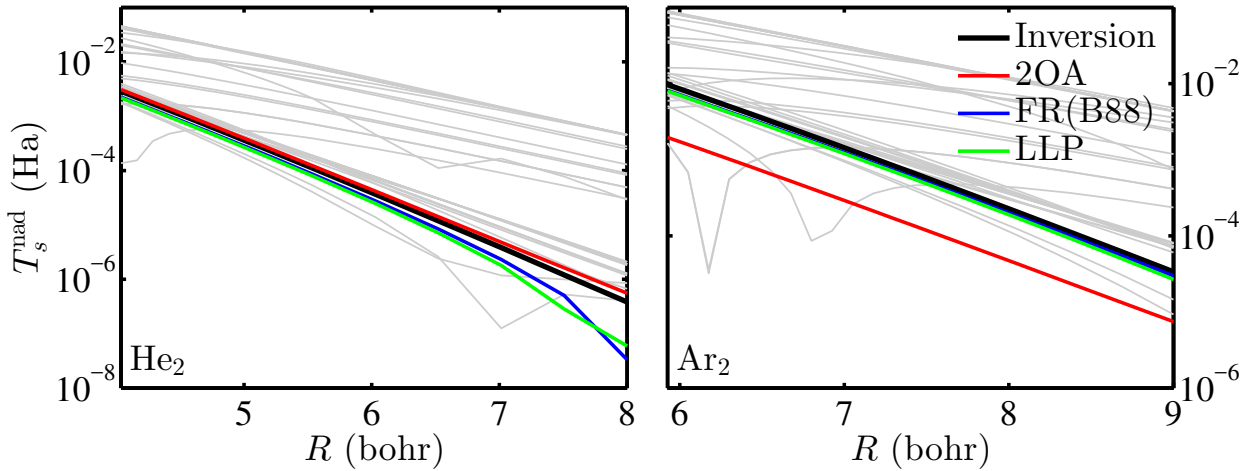}
	\caption{The approximate NAKE {\em vs.} the inter-nuclear separation for the rare-gas dimers. The curves in gray correspond to approximate functionals in Libxc \cite{MOB12}, all of which fail to reproduce the exact asymptotic behavior. The \textquotedblleft{best}\textquotedblright ones, \FRBee and \LLP, are highlighted alongside the 2OA.}
	\label{fig:2oa_asympt}
\end{figure}

As we noted in section \ref{sec:asympt_exact}, the exact NAKE behaves as a nearly exponential function of the inter-nuclear separation for all the rare-gas dimers. The 2OA of Eq.(\ref{eqn:2oa}) satisfies this exact condition, as shown numerically in  
Figure \ref{fig:2oa_asympt}, where the y-axis is in the logarithmic scale. 
The 2OA not only reproduces the nearly exponential behavior but also has the same decay constant compared to the exact NAKE. 
On the other hand, most of the other approximations either do not reproduce the nearly exponential behavior or reproduce the nearly exponential behavior but with a different decay constant (\FRBee and \LLP reproduce the asymptotic behavior, but their decay constants are not as accurate as 2OA).

\begin{figure}[htb]
	\includegraphics*[width=4in]{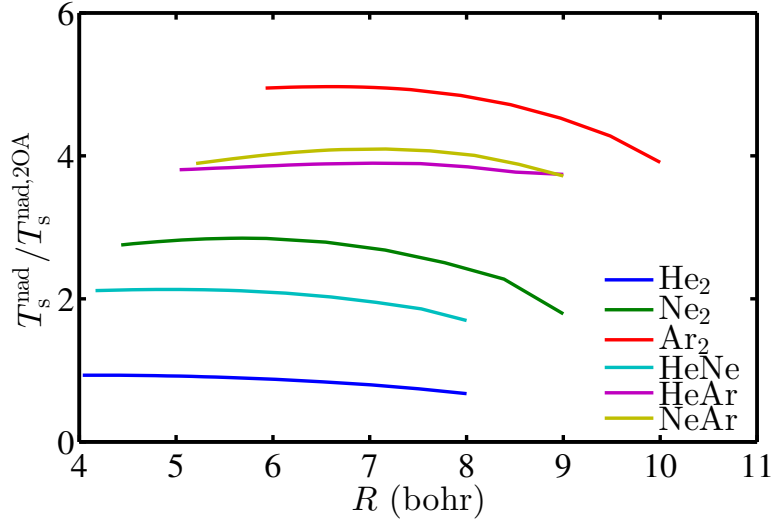}
	\caption{The ratio between the NAKE from inversion and 2OA vs the inter-nuclear separation for the rare-gas dimers.}
	\label{fig:2oa_magic}
\end{figure}

Figure \ref{fig:2oa_magic} compares the ratio between the exact NAKE and the 2OA.
For each system, the ratio is almost a constant regardless of the inter-nuclear separation. 
This indicates that we can multiply 2OA by a single system-dependent parameter $M$ and obtain an approximation that accurately reproduces the NAKE for all rare-gas dimers. 
We refer to 2OA with the parameter $M$ as scaled-2OA here onwards. 
For He$_2$, we expect $M$ to be slightly less than 1 because the KE follows a variational principle and therefore the correct KE for the two orbitals of the helium dimer must be less than the KE of our approximate orbitals.

We optimized $M$ by minimizing the square of the difference between the NAKE from scaled-2OA and inversion. 
Only the data between ${R\sss{eq}}-0.5 \mathrm{bohr}$ and ${R\sss{eq}}+0.5 \mathrm{bohr}$ are used for the minimization, where $R\sss{eq}$ is the equilibrium separation. 
The results are shown in Table \ref{tab:magic_number}.

\begin{table*}[htb]
	\begin{tabular}{l S[table-format=1.2]}
		\toprule
		{System} & {$M$}   \\ 
		\midrule
		He$_2$ & 0.93 \\
		Ne$_2$ & 2.80 \\
		Ar$_2$ & 4.96 \\
		HeNe   & 2.12 \\
		HeAr   & 3.83 \\
		NeAr   & 3.96 \\
		\bottomrule		
	\end{tabular}
	\caption{The optimized $M$ for the rare-gas dimers.}
	\label{tab:magic_number}
\end{table*}

\subsection{Dissociation}

\begin{figure}[htb]
	\includegraphics*[width=\textwidth]{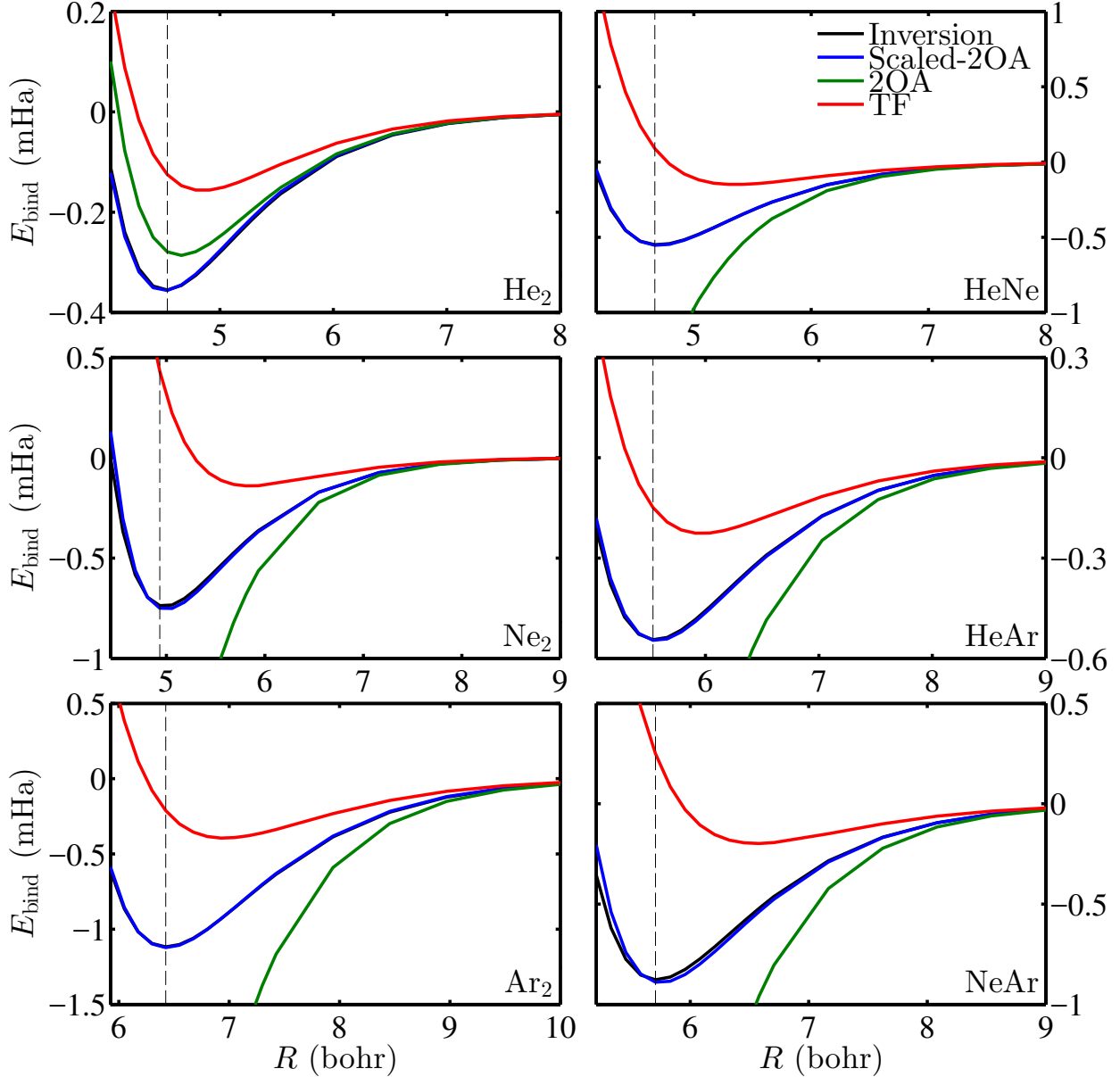}
	\caption{Binding Curves for rare-gas dimers with 2OA, scaled-2OA and TF. $R$ is the inter-nuclear separation and $E_\mathrm{bind}$ is the binding energy. The curves of 2OA and scaled-2OA are from non-self-consistent P-DFT calculations. Dashed lines indicate the equilibrium separation.}
	\label{fig:binding_2oa}
\end{figure}

Figure \ref{fig:binding_2oa} compares the binding curve for inversion, TF, 2OA and scaled-2OA. 
As expected, the 2OA only works well in the case of He$_2$. 
However, it is impressive that the scaled-2OA binding curves accurately match the exact binding curves in {\em all} these cases.

\section{Conclusions}

We have provided uniquely defined numerically exact NAKE reference data as well as NAKE data from approximate KE functionals for the rare-gas dimers.
We also introduced two new NAKE functional approximations: \Repar, a conjoint PBE functional re-parameterized with the NAKE data for the rare-gas dimers; and 2OA, a simple, physically-motivated, non-decomposable NAKE approximation constructed from two orbitals that are explicit functionals of the fragment densities.

Our NAKE data obtained from highly accurate inversion procedure with P-DFT calculation show that the exact NAKE for the rare-gas dimers has a nearly exponential asymptotic behavior, which is not well reproduced by the approximate KE functionals. 
Our new approximations significantly improve this asymptotic behavior. 
For \Repar, this is a result of fitting into the NAKE data. 
Although 2OA was not intended to approximate systems with more than two orbitals, it also reproduces the asymptotic behavior extremely well for all the rare-gas dimers. 

We also provided the data of the NAKE per particle and the NAKP, two quantities that can be used to measure the performance of NAKE approximations in different regions. 
None of the approximations matches the exact result accurately for these quantities. 
However, \Repar can reproduce the main feature of a double peak and a central well, but the size of these features is incorrect at large separations.

As for the NAKE and the self-consistent density at the equilibrium,
approximations that use the conjointness-conjecture strategy generally produce better results while approximations based on the linear combinations of \TFa and \VWa functionals produce worse results even than setting \none. 
For the NAKE of rare-gas dimers, \Repar provides the maximum accuracy attainable from conjoint PBE functionals.

Further improvement of approximate functionals for $T\s\nad$ remains an important open challenge.

\section{Acknowledgments}
We acknowledge support from the Office of Basic Energy Sciences, US Department of Basic Energy Sciences, US Department of Energy (DOE), under grant DE-FG02-10ER16191. 
A.W. also acknowledges support from the US National Science Foundation CAREER program under grant CHE-1149968, and from the Camille Dreyfus Teacher-Scholar Awards Program.

\bibliographystyle{naturemag}
\bibliography{JNW16}

\section{Supplemental Data}
\begin{longtable}{ l *{6}{S[detect-all, table-format=1.1, table-figures-exponent = 1, table-auto-round, scientific-notation=true]}}\\
	\toprule
	{Functional} & {He$_2$} & {Ne$_2$} & {Ar$_2$} & {HeNe} & {HeAr} & {NeAr}\\
	\midrule
	\none       & -0.000992935 & -0.002749803 & -0.004050443 & -0.001844909 & -0.001802383 & -0.003151847 \\
	\TFa        & 0.000204941  & 0.001527064  & 0.000724964  & 0.001057245  & 0.000368284  & 0.001054373  \\
	\VWa        & \itshape -0.038816314 & \itshape -0.078574121 & \itshape -0.081391498 & \itshape -0.058621124 & \itshape -0.053375019 & \itshape -0.079174517 \\
	\TFVW       & -0.034479013 & -0.065908375 & -0.066487650 & -0.050343185 & -0.045722064 & -0.065203056 \\
	\GEt        & -0.002553623 & -0.003884723 & -0.004806448 & -0.003068981 & -0.003321866 & -0.004245725 \\
	\GOLDEN     & -0.007525784 & -0.013624433 & -0.014685987 & -0.010490816 & -0.009964738 & -0.013795306 \\
	\YTsf       & -0.004953120 & -0.008587806 & -0.009588490 & -0.006653262 & -0.006529146 & -0.008855465 \\
	\BALTIN     & -0.016297154 & -0.030739974 & -0.031847470 & -0.023530293 & -0.021649730 & -0.030601214 \\
	\LIEB       & -0.004561240 & -0.007820045 & -0.008809344 & -0.006068212 & -0.006005526 & -0.008102734 \\
	\Repar      & \bfseries 0.000002206  & 0.000626711  & 0.000260030  & 0.000255579  & \bfseries 0.000081079  & 0.000625061  \\
	\REVAPBEINT & -0.000050265 & 0.000639392  & -0.000654651 & 0.000492594  & -0.000174947 & 0.000063049  \\
	\APBEINT    & 0.000135257  & 0.001148103  & -0.000023765 & 0.000841082  & 0.000143379  & 0.000590913  \\
	\REVAPBE    & -0.000024458 & 0.000714198  & -0.000562756 & 0.000542892  & -0.000128850 & 0.000142425  \\
	\APBE       & 0.000155096  & 0.001209034  & 0.000056365  & 0.000881213  & 0.000180714  & 0.000657349  \\
	\TWa        & 0.000142851  & 0.001177547  & 0.000020797  & 0.000859114  & 0.000160836  & 0.000625420  \\
	\TWb        & 0.000200493  & 0.001346924  & 0.000246111  & 0.000972683  & 0.000266217  & 0.000805601  \\
	\TWc        & 0.000132130  & 0.001147898  & \bfseries -0.000016151 & 0.000838838  & 0.000142280  & 0.000594538  \\
	\TWd        & 0.000125063  & 0.001128482  & -0.000040188 & 0.000825533  & 0.000130121  & 0.000574363  \\
	\LLP        & -0.000236211 & 0.000439748  & -0.000650767 & 0.000291545  & -0.000335285 & -0.000072914 \\
	\FRBee      & -0.000205452 & 0.000510921  & -0.000571979 & 0.000342903  & -0.000289167 & \bfseries -0.000001301 \\
	\THAKKAR    & -0.000376362 & 0.000074175  & -0.001111929 & 0.000040308  & -0.000565121 & -0.000457293 \\
	\LCnr       & -0.000427578 & \bfseries 0.000004412  & -0.001124647 & \bfseries -0.000024549 & -0.000617789 & -0.000507048 \\
	\FRPWes     & 0.000066071  & 0.001098171  & 0.000038088  & 0.000779022  & 0.000099463  & 0.000575098  \\
	\LP\cite{LP87}         & 0.000243319  & 0.001646576  & 0.000871634  & 0.001133815  & 0.000436595  & 0.001185120  \\
	\OLa\cite{OL91}        & -0.001513808 & -0.001848345 & -0.002735228 & -0.001515842 & -0.001933152 & -0.002251680 \\
	\OLb\cite{OL91}        & -0.002553007 & -0.003882853 & -0.004804194 & -0.003067772 & -0.003320783 & -0.004243690 \\
	\PEARSON\cite{PG85}    & 0.000297262  & 0.001817533  & 0.001138983  & 0.001248244  & 0.000551569  & 0.001385099  \\
	\PERDEW\cite{Per92}     & 0.000357767  & 0.002054799  & 0.001562676  & 0.001395745  & 0.000695717  & 0.001646070  \\
	\VSK\cite{VSK98}        & -0.029118519 & -0.050248749 & -0.047870005 & -0.039654806 & -0.036289899 & -0.048765132 \\
	\VJKS\cite{VJKS00}       & 0.010134562  & 0.019342644  & 0.016989641  & 0.015137572  & 0.012797627  & 0.017979914  \\
	\ERNZERHOF\cite{Ern00}  & -0.012599464 & -0.019566938 & -0.018318737 & -0.016098078 & -0.014952540 & -0.019007876 \\
	\bottomrule
	\caption{The error in the NAKE of various KE functionals. For each system, the best result is in boldface and the worst is in italics.}
	\label{tab:err_tsnad_all}
\end{longtable}

\begin{longtable}{ l *{6}{S[detect-all, table-format=1.1, table-figures-exponent = 1, table-auto-round, scientific-notation=true]}}\\
	\toprule
	{Functional} & {He$_2$} & {Ne$_2$} & {Ar$_2$} & {HeNe} & {HeAr} & {NeAr}\\
	\midrule
	\none       & 0.00534874 & 0.01063406 & 0.02142618 & 0.00849323 & 0.01125317 & 0.01452504 \\
	\TFa        & 0.00173069 & 0.00456821 & 0.00735248 & 0.00304807 & 0.00388457 & 0.00558493 \\
	\VWa        & \itshape 0.13640329 & \itshape 0.23635308 & \itshape 0.32698531 & \itshape 0.18914528 & \itshape 0.21060381 & \itshape 0.27557036 \\
	\TFVW       & 0.12071070 & 0.20071513 & 0.26943977 & 0.16305288 & 0.18107313 & 0.23134484 \\
	\GEt        & 0.01099064 & 0.01639058 & 0.02328452 & 0.01420872 & 0.01723691 & 0.01945147 \\
	\GOLDEN     & 0.03082387 & 0.05033483 & 0.06947932 & 0.04144438 & 0.04725125 & 0.05866698 \\
	\YTsf       & 0.02075976 & 0.03314817 & 0.04611335 & 0.02763653 & 0.03203447 & 0.03880368 \\
	\BALTIN     & 0.06265744 & 0.10422954 & 0.14204880 & 0.08486004 & 0.09509222 & 0.12077756 \\
	\LIEB       & 0.01919105 & 0.03046163 & 0.04245204 & 0.02548085 & 0.02965745 & 0.03569860 \\
	\Repar      & 0.00151018 & 0.00298875 & \bfseries 0.00604696 & 0.00226527 & \bfseries 0.00376447 & \bfseries 0.00445767 \\
	\REVAPBEINT & 0.00296497 & 0.00579438 & 0.01170741 & 0.00473871 & 0.00734588 & 0.00863822 \\
	\APBEINT    & 0.00177992 & 0.00308356 & 0.00792183 & 0.00272343 & 0.00512671 & 0.00581736 \\
	\REVAPBE    & 0.00281139 & 0.00540965 & 0.01119282 & 0.00446332 & 0.00706180 & 0.00826670 \\
	\APBE       & 0.00169300 & 0.00285330 & 0.00749173 & 0.00255087 & 0.00492354 & 0.00553045 \\
	\TWa        & 0.00174258 & 0.00293904 & 0.00759664 & 0.00262676 & 0.00500733 & 0.00562789 \\
	\TWb        & \bfseries 0.00145374 & \bfseries 0.00231568 & 0.00640231 & \bfseries 0.00210744 & 0.00439581 & 0.00472786 \\
	\TWc        & 0.00179784 & 0.00305646 & 0.00777804 & 0.00272134 & 0.00511850 & 0.00577125 \\
	\TWd        & 0.00183482 & 0.00313506 & 0.00789786 & 0.00278474 & 0.00519214 & 0.00586431 \\
	\LLP        & 0.00277454 & 0.00405821 & 0.00840911 & 0.00378736 & 0.00618996 & 0.00651277 \\
	\FRBee      & 0.00262848 & 0.00377747 & 0.00806487 & 0.00357004 & 0.00595793 & 0.00624929 \\
	\THAKKAR    & 0.00370282 & 0.00619357 & 0.01177454 & 0.00532307 & 0.00787766 & 0.00880657 \\
	\LCnr       & 0.00375590 & 0.00586362 & 0.01068964 & 0.00519749 & 0.00766296 & 0.00826116 \\
	\FRPWes     & 0.00167458 & 0.00235703 & 0.00642782 & 0.00226711 & 0.00451617 & 0.00478404 \\
	\LP         & 0.00187532 & 0.00498458 & 0.00789105 & 0.00329522 & 0.00401239 & 0.00600155 \\
	\OLa        & 0.00667141 & 0.00904853 & 0.01338763 & 0.00829898 & 0.01076831 & 0.01104177 \\
	\OLb        & 0.01098772 & 0.01638393 & 0.02327322 & 0.01420381 & 0.01723144 & 0.01944315 \\
	\PEARSON    & 0.00242263 & 0.00702140 & 0.01221118 & 0.00453760 & 0.00481484 & 0.00864376 \\
	\PERDEW     & 0.00397768 & 0.01089853 & 0.01753191 & 0.00721450 & 0.00729420 & 0.01323043 \\
	\VSK        & 0.07158185 & 0.09056849 & 0.10998364 & 0.08225605 & 0.09089340 & 0.10005833 \\
	\VJKS       & 0.03416226 & 0.04692909 & 0.05269517 & 0.04091206 & 0.04237928 & 0.04990850 \\
	\ERNZERHOF  & 0.03396301 & 0.04398123 & 0.05639833 & 0.03981859 & 0.04420530 & 0.04832510 \\
	\bottomrule
	\caption{The error in the self-consistent density of various KE functionals. For each system, the best result is in boldface and the worst is in italics.}
	\label{tab:err_den_all}
\end{longtable}

\end{document}

%% file: mydefs.tex
\def\bea{\begin{eqnarray}}
\def\eea{\end{eqnarray}}
\def\ben{\begin{equation}}
\def\een{\end{equation}}
\def\benu{\begin{enumerate}}
\def\enu{\end{enumerate}}
\def\beal{\begin{equation}\begin{aligned}} 
\def\eeal{\end{aligned}\end{equation}}

\def\sss{\scriptscriptstyle\rm}

\def\half{\frac{1}{2}}

\def\br{{\mathbf r}}

\def\d3{{\rm d^3}}
\def\intdhr{\int\d3{r}}

\def\s{_{\sss S}}
\def\p{_{\sss p}}

\def\xc{_{\sss XC}}

\def\H{_{\sss H}}

\def\ext{_{\rm ext}}

\def\TF{^{\rm TF}}

\def\PBE{^{\rm PBE}}

\def\VW{^{\rm vW}}

\def\nad{^{\rm nad}}

\usepackage{xspace}

\newcommand{\none}{$T\s\nad=0$\xspace}
\newcommand{\Repar}{R-PBE\xspace}
\newcommand{\TFa}{TF\xspace}
\newcommand{\LP}{LP\xspace}
\newcommand{\TFVW}{TFvW\xspace}
\newcommand{\REVAPBEINT}{revAPBEKint\xspace}
\newcommand{\APBEINT}{APBEKint\xspace}
\newcommand{\REVAPBE}{revAPBEK\xspace}
\newcommand{\APBE}{APBEK\xspace}
\newcommand{\TWa}{TW02I\xspace}
\newcommand{\TWb}{TW02II\xspace}
\newcommand{\TWc}{TW02III\xspace}
\newcommand{\TWd}{TW02IV\xspace}
\newcommand{\VWa}{vW\xspace}
\newcommand{\GEt}{GEA2\xspace}
\newcommand{\GOLDEN}{Golden\xspace}
\newcommand{\YTsf}{YT65\xspace}
\newcommand{\BALTIN}{Baltin\xspace}
\newcommand{\LIEB}{Lieb\xspace}

\newcommand{\PEARSON}{P82\xspace}
\newcommand{\OLa}{OL1\xspace}
\newcommand{\OLb}{OL2\xspace}
\newcommand{\FRBee}{FR(B88)\xspace}
\newcommand{\FRPWes}{FR(PW86)\xspace}
\newcommand{\PERDEW}{P92\xspace}
\newcommand{\VSK}{VSK98\xspace}
\newcommand{\VJKS}{VJKS00\xspace}
\newcommand{\ERNZERHOF}{E00\xspace}
\newcommand{\LCnr}{LC94\xspace}
\newcommand{\LLP}{LLP\xspace}
\newcommand{\THAKKAR}{T92\xspace}
\newcommand{\mTFVW}{TF$\lambda$W\xspace}

\def\Tsnad{T\s\nad[\{n_\alpha\}]}
\def\epkin{t\s\nad}
\def\vpkin{v_{\sss t}\nad}

%% file: JNW16.bbl
\begin{thebibliography}{10}
\expandafter\ifx\csname url\endcsname\relax
  \def\url#1{\texttt{#1}}\fi
\expandafter\ifx\csname urlprefix\endcsname\relax\def\urlprefix{URL }\fi
\providecommand{\bibinfo}[2]{#2}
\providecommand{\eprint}[2][]{\url{#2}}

\bibitem{Cor91}
\bibinfo{author}{Cortona, P.}
\newblock \bibinfo{title}{Self-consistently determined properties of solids
  without band-structure calculations}.
\newblock \emph{\bibinfo{journal}{Physical Review B}}
  \textbf{\bibinfo{volume}{44}}, \bibinfo{pages}{8454} (\bibinfo{year}{1991}).

\bibitem{WW93}
\bibinfo{author}{Weso{\l}owski, T.~A.} \& \bibinfo{author}{Warshel, A.}
\newblock \bibinfo{title}{Frozen density functional approach for ab initio
  calculations of solvated molecules}.
\newblock \emph{\bibinfo{journal}{The Journal of Physical Chemistry}}
  \textbf{\bibinfo{volume}{97}}, \bibinfo{pages}{8050--8053}
  (\bibinfo{year}{1993}).

\bibitem{JN14}
\bibinfo{author}{Jacob, C.~R.} \& \bibinfo{author}{Neugebauer, J.}
\newblock \bibinfo{title}{Subsystem density-functional theory}.
\newblock \emph{\bibinfo{journal}{Wiley Interdisciplinary Reviews:
  Computational Molecular Science}} \textbf{\bibinfo{volume}{4}},
  \bibinfo{pages}{325--362} (\bibinfo{year}{2014}).

\bibitem{FJN+10}
\bibinfo{author}{Fux, S.}, \bibinfo{author}{Jacob, C.~R.},
  \bibinfo{author}{Neugebauer, J.}, \bibinfo{author}{Visscher, L.} \&
  \bibinfo{author}{Reiher, M.}
\newblock \bibinfo{title}{Accurate frozen-density embedding potentials as a
  first step towards a subsystem description of covalent bonds}.
\newblock \emph{\bibinfo{journal}{The Journal of Chemical Physics}}
  \textbf{\bibinfo{volume}{132}}, \bibinfo{pages}{164101}
  (\bibinfo{year}{2010}).

\bibitem{GAMMI10}
\bibinfo{author}{Goodpaster, J.~D.}, \bibinfo{author}{Ananth, N.},
  \bibinfo{author}{Manby, F.~R.} \& \bibinfo{author}{Miller~III, T.~F.}
\newblock \bibinfo{title}{Exact nonadditive kinetic potentials for embedded
  density functional theory}.
\newblock \emph{\bibinfo{journal}{The Journal of Chemical Physics}}
  \textbf{\bibinfo{volume}{133}}, \bibinfo{pages}{084103}
  (\bibinfo{year}{2010}).

\bibitem{HPC11}
\bibinfo{author}{Huang, C.}, \bibinfo{author}{Pavone, M.} \&
  \bibinfo{author}{Carter, E.~A.}
\newblock \bibinfo{title}{Quantum mechanical embedding theory based on a unique
  embedding potential}.
\newblock \emph{\bibinfo{journal}{The Journal of Chemical Physics}}
  \textbf{\bibinfo{volume}{134}}, \bibinfo{pages}{154110}
  (\bibinfo{year}{2011}).

\bibitem{NWW11}
\bibinfo{author}{Nafziger, J.}, \bibinfo{author}{Wu, Q.} \&
  \bibinfo{author}{Wasserman, A.}
\newblock \bibinfo{title}{Molecular binding energies from partition density
  functional theory}.
\newblock \emph{\bibinfo{journal}{The Journal of Chemical Physics}}
  \textbf{\bibinfo{volume}{135}}, \bibinfo{pages}{234101}
  (\bibinfo{year}{2011}).

\bibitem{NJW17}
\bibinfo{author}{Nafziger, J.}, \bibinfo{author}{Jiang, K.} \&
  \bibinfo{author}{Wasserman, A.}
\newblock \bibinfo{title}{Accurate reference data for the nonadditive,
  noninteracting kinetic energy in covalent bonds}.
\newblock \emph{\bibinfo{journal}{Journal of Chemical Theory and Computation}}
  \textbf{\bibinfo{volume}{13}}, \bibinfo{pages}{577--586}
  (\bibinfo{year}{2017}).

\bibitem{JW17}
\bibinfo{author}{Jensen, D.~S.} \& \bibinfo{author}{Wasserman, A.}
\newblock \bibinfo{title}{Numerical methods for the inverse problem of density
  functional theory}.
\newblock \emph{\bibinfo{journal}{International Journal of Quantum Chemistry}}
  \bibinfo{pages}{e25425--n/a} (\bibinfo{year}{2017}).
\newblock \bibinfo{note}{E25425}.

\bibitem{CW07}
\bibinfo{author}{Cohen, M.~H.} \& \bibinfo{author}{Wasserman, A.}
\newblock \bibinfo{title}{On the {Foundations} of {Chemical Reactivity
  Theory}}.
\newblock \emph{\bibinfo{journal}{The Journal of Physical Chemistry A}}
  \textbf{\bibinfo{volume}{111}}, \bibinfo{pages}{2229--2242}
  (\bibinfo{year}{2007}).

\bibitem{EBCW10}
\bibinfo{author}{Elliott, P.}, \bibinfo{author}{Burke, K.},
  \bibinfo{author}{Cohen, M.~H.} \& \bibinfo{author}{Wasserman, A.}
\newblock \bibinfo{title}{Partition density-functional theory}.
\newblock \emph{\bibinfo{journal}{Physical Review A}}
  \textbf{\bibinfo{volume}{82}}, \bibinfo{pages}{024501}
  (\bibinfo{year}{2010}).

\bibitem{NW14}
\bibinfo{author}{Nafziger, J.} \& \bibinfo{author}{Wasserman, A.}
\newblock \bibinfo{title}{Density-{Based Partitioning Methods} for
  {Ground-State Molecular Calculations}}.
\newblock \emph{\bibinfo{journal}{The Journal of Physical Chemistry A}}
  \textbf{\bibinfo{volume}{118}}, \bibinfo{pages}{7623--7639}
  (\bibinfo{year}{2014}).

\bibitem{Tho26}
\bibinfo{author}{Thomas, L.~H.}
\newblock \bibinfo{title}{The calculation of atomic fields}.
\newblock \emph{\bibinfo{journal}{Mathematical Proceedings of the Cambridge
  Philosophical Society}} \textbf{\bibinfo{volume}{23}},
  \bibinfo{pages}{542--548} (\bibinfo{year}{1926}).

\bibitem{Fer27}
\bibinfo{author}{Fermi, E.}
\newblock \bibinfo{title}{Un metodo statistico per la determinazione di alcune
  priorieta dell'atome}.
\newblock \emph{\bibinfo{journal}{Rendiconti. Accademia Nazionale dei Lincei}}
  \textbf{\bibinfo{volume}{6}}, \bibinfo{pages}{32} (\bibinfo{year}{1927}).

\bibitem{Wei35}
\bibinfo{author}{v~Weizs{\"a}cker, C.~F.}
\newblock \bibinfo{title}{Zur theorie der kernmassen}.
\newblock \emph{\bibinfo{journal}{Zeitschrift F{\"u}r Physik a Hadrons and
  Nuclei}} \textbf{\bibinfo{volume}{96}}, \bibinfo{pages}{431--458}
  (\bibinfo{year}{1935}).

\bibitem{TW13}
\bibinfo{author}{Tran, F.} \& \bibinfo{author}{Weso{\l}owski, T.~A.}
\newblock \bibinfo{title}{Semilocal approximations for the kinetic energy}.
\newblock \emph{\bibinfo{journal}{Recent Progress in Orbital-free Density
  Functional Theory}} \bibinfo{pages}{429--442} (\bibinfo{year}{2013}).

\bibitem{KP56}
\bibinfo{author}{Kompaneets, A.~S.} \& \bibinfo{author}{Pavlovsky, E.~S.}
\newblock \bibinfo{title}{Self-consistent equations for atoms}.
\newblock \emph{\bibinfo{journal}{Journal of Experimental and Theoretical
  Physics}} \textbf{\bibinfo{volume}{31}}, \bibinfo{pages}{427}
  (\bibinfo{year}{1956}).

\bibitem{Kir57a}
\bibinfo{author}{Kirzhnits, D.~A.}
\newblock \bibinfo{title}{Quantum corrections to the thomas-fermi equation}.
\newblock \emph{\bibinfo{journal}{Journal of Experimental and Theoretical
  Physics}} \textbf{\bibinfo{volume}{5}} (\bibinfo{year}{1957}).

\bibitem{Gol57}
\bibinfo{author}{Golden, S.}
\newblock \bibinfo{title}{Statistical theory of many-electron systems. general
  considerations pertaining to the thomas-fermi theory}.
\newblock \emph{\bibinfo{journal}{Physical Review}}
  \textbf{\bibinfo{volume}{105}}, \bibinfo{pages}{604--615}
  (\bibinfo{year}{1957}).

\bibitem{YT65}
\bibinfo{author}{Yonei, K.} \& \bibinfo{author}{Tomishima, Y.}
\newblock \bibinfo{title}{On the weizs{\"{a}}cker correction to the
  {Thomas}-{Fermi} theory of the atom}.
\newblock \emph{\bibinfo{journal}{Journal of the Physical Society of Japan}}
  \textbf{\bibinfo{volume}{20}}, \bibinfo{pages}{1051--1057}
  (\bibinfo{year}{1965}).

\bibitem{Bal72}
\bibinfo{author}{Baltin, R.}
\newblock \bibinfo{title}{The energy-density functional of an electron gas in
  locally linear approximation of the one-body potential}.
\newblock \emph{\bibinfo{journal}{Zeitschrift F{\"u}r Naturforschung A}}
  \textbf{\bibinfo{volume}{27}}, \bibinfo{pages}{1176--1186}
  (\bibinfo{year}{1972}).

\bibitem{Lie81}
\bibinfo{author}{Lieb, E.~H.}
\newblock \bibinfo{title}{Thomas-fermi and related theories of atoms and
  molecules}.
\newblock \emph{\bibinfo{journal}{Reviews of Modern Physics}}
  \textbf{\bibinfo{volume}{53}}, \bibinfo{pages}{603--641}
  (\bibinfo{year}{1981}).

\bibitem{MS91}
\bibinfo{author}{March, N.~H.} \& \bibinfo{author}{Santamaria, R.}
\newblock \bibinfo{title}{Non-local relation between kinetic and exchange
  energy densities in hartree--fock theory}.
\newblock \emph{\bibinfo{journal}{International Journal of Quantum Chemistry}}
  \textbf{\bibinfo{volume}{39}}, \bibinfo{pages}{585--592}
  (\bibinfo{year}{1991}).

\bibitem{LLP91}
\bibinfo{author}{Lee, H.}, \bibinfo{author}{Lee, C.} \& \bibinfo{author}{Parr,
  R.~G.}
\newblock \bibinfo{title}{Conjoint gradient correction to the {Hartree}-{Fock}
  kinetic- and exchange-energy density functionals}.
\newblock \emph{\bibinfo{journal}{Physical Review A}}
  \textbf{\bibinfo{volume}{44}}, \bibinfo{pages}{768--771}
  (\bibinfo{year}{1991}).

\bibitem{PBE96}
\bibinfo{author}{Perdew, J.~P.}, \bibinfo{author}{Burke, K.} \&
  \bibinfo{author}{Ernzerhof, M.}
\newblock \bibinfo{title}{Generalized {Gradient Approximation Made Simple}}.
\newblock \emph{\bibinfo{journal}{Physical Review Letters}}
  \textbf{\bibinfo{volume}{77}}, \bibinfo{pages}{3865--3868}
  (\bibinfo{year}{1996}).

\bibitem{TW02a}
\bibinfo{author}{Tran, F.} \& \bibinfo{author}{Weso{\l}owski, T.~A.}
\newblock \bibinfo{title}{Link between the kinetic- and exchange-energy
  functionals in the generalized gradient approximation}.
\newblock \emph{\bibinfo{journal}{International Journal of Quantum Chemistry}}
  \textbf{\bibinfo{volume}{89}}, \bibinfo{pages}{441--446}
  (\bibinfo{year}{2002}).

\bibitem{CFLDS11}
\bibinfo{author}{Constantin, L.~A.}, \bibinfo{author}{Fabiano, E.},
  \bibinfo{author}{Laricchia, S.} \& \bibinfo{author}{Della~Sala, F.}
\newblock \bibinfo{title}{Semiclassical {Neutral Atom} as a {Reference System}
  in {Density Functional Theory}}.
\newblock \emph{\bibinfo{journal}{Physical Review Letters}}
  \textbf{\bibinfo{volume}{106}}, \bibinfo{pages}{186406}
  (\bibinfo{year}{2011}).

\bibitem{LFCDS11}
\bibinfo{author}{Laricchia, S.}, \bibinfo{author}{Fabiano, E.},
  \bibinfo{author}{Constantin, L.~A.} \& \bibinfo{author}{Della~Sala, F.}
\newblock \bibinfo{title}{Generalized {Gradient Approximations} of the
  {Noninteracting Kinetic Energy} from the {Semiclassical Atom Theory}:
  {Rationalization} of the {Accuracy} of the {Frozen Density Embedding Theory}
  for {Nonbonded Interactions}}.
\newblock \emph{\bibinfo{journal}{Journal of Chemical Theory and Computation}}
  \textbf{\bibinfo{volume}{7}}, \bibinfo{pages}{2439--2451}
  (\bibinfo{year}{2011}).

\bibitem{FR95}
\bibinfo{author}{Fuentealba, P.} \& \bibinfo{author}{Reyes, O.}
\newblock \bibinfo{title}{Further evidence of the conjoint correction to the
  local kinetic and exchange energy density functionals}.
\newblock \emph{\bibinfo{journal}{Chemical Physics Letters}}
  \textbf{\bibinfo{volume}{232}}, \bibinfo{pages}{31--34}
  (\bibinfo{year}{1995}).

\bibitem{LC94}
\bibinfo{author}{Lembarki, A.} \& \bibinfo{author}{Chermette, H.}
\newblock \bibinfo{title}{Obtaining a gradient-corrected kinetic-energy
  functional from the {Perdew-Wang} exchange functional}.
\newblock \emph{\bibinfo{journal}{Physical Review A}}
  \textbf{\bibinfo{volume}{50}}, \bibinfo{pages}{5328--5331}
  (\bibinfo{year}{1994}).

\bibitem{Tha92}
\bibinfo{author}{Thakkar, A.~J.}
\newblock \bibinfo{title}{Comparison of kinetic-energy density functionals}.
\newblock \emph{\bibinfo{journal}{Physical Review A}}
  \textbf{\bibinfo{volume}{46}}, \bibinfo{pages}{6920--6924}
  (\bibinfo{year}{1992}).

\bibitem{WGC99}
\bibinfo{author}{Wang, Y.~A.}, \bibinfo{author}{Govind, N.} \&
  \bibinfo{author}{Carter, E.~A.}
\newblock \bibinfo{title}{Orbital-free kinetic-energy density functionals with
  a density-dependent kernel}.
\newblock \emph{\bibinfo{journal}{Physical Review B}}
  \textbf{\bibinfo{volume}{60}}, \bibinfo{pages}{16350--16358}
  (\bibinfo{year}{1999}).

\bibitem{HC10}
\bibinfo{author}{Huang, C.} \& \bibinfo{author}{Carter, E.~A.}
\newblock \bibinfo{title}{Nonlocal orbital-free kinetic energy density
  functional for semiconductors}.
\newblock \emph{\bibinfo{journal}{Physical Review B}}
  \textbf{\bibinfo{volume}{81}}, \bibinfo{pages}{045206}
  (\bibinfo{year}{2010}).

\bibitem{XC15}
\bibinfo{author}{Xia, J.} \& \bibinfo{author}{Carter, E.~A.}
\newblock \bibinfo{title}{Single-point kinetic energy density functionals: {A}
  pointwise kinetic energy density analysis and numerical convergence
  investigation}.
\newblock \emph{\bibinfo{journal}{Physical Review B}}
  \textbf{\bibinfo{volume}{91}}, \bibinfo{pages}{045124}
  (\bibinfo{year}{2015}).

\bibitem{KT15}
\bibinfo{author}{Karasiev, V.~V.} \& \bibinfo{author}{Trickey, S.~B.}
\newblock \bibinfo{title}{Chapter nine-frank discussion of the status of
  ground-state orbital-free dft}.
\newblock \emph{\bibinfo{journal}{Advances in Quantum Chemistry}}
  \textbf{\bibinfo{volume}{71}}, \bibinfo{pages}{221--245}
  (\bibinfo{year}{2015}).

\bibitem{CFDS17}
\bibinfo{author}{Constantin, L.~A.}, \bibinfo{author}{Fabiano, E.} \&
  \bibinfo{author}{Della~Sala, F.}
\newblock \bibinfo{title}{Modified fourth-order kinetic energy gradient
  expansion with hartree potential-dependent coefficients}.
\newblock \emph{\bibinfo{journal}{Journal of Chemical Theory and Computation}}
  \textbf{\bibinfo{volume}{13}}, \bibinfo{pages}{4228--4239}
  (\bibinfo{year}{2017}).

\bibitem{GP17}
\bibinfo{author}{Genova, A.} \& \bibinfo{author}{Pavanello, M.}
\newblock \bibinfo{title}{Nonlocal kinetic energy functionals by functional
  integration}.
\newblock \emph{\bibinfo{journal}{Arxiv Preprint Arxiv:1704.08943}}
  (\bibinfo{year}{2017}).

\bibitem{KH15}
\bibinfo{author}{Koch, W.} \& \bibinfo{author}{Holthausen, M.~C.}
\newblock \emph{\bibinfo{title}{A chemist's guide to density functional
  theory}} (\bibinfo{publisher}{John Wiley \& Sons}, \bibinfo{year}{2015}).

\bibitem{WNJ+17}
\bibinfo{author}{Wasserman, A.} \emph{et~al.}
\newblock \bibinfo{title}{The importance of being inconsistent}.
\newblock \emph{\bibinfo{journal}{Annual Review of Physical Chemistry}}
  \textbf{\bibinfo{volume}{68}}, \bibinfo{pages}{555--581}
  (\bibinfo{year}{2017}).

\bibitem{VBG+10}
\bibinfo{author}{Valiev, M.} \emph{et~al.}
\newblock \bibinfo{title}{Nwchem: A comprehensive and scalable open-source
  solution for large scale molecular simulations}.
\newblock \emph{\bibinfo{journal}{Computer Physics Communications}}
  \textbf{\bibinfo{volume}{181}}, \bibinfo{pages}{1477--1489}
  (\bibinfo{year}{2010}).

\bibitem{SLBB03}
\bibinfo{author}{Sim, E.}, \bibinfo{author}{Larkin, J.},
  \bibinfo{author}{Burke, K.} \& \bibinfo{author}{Bock, C.~W.}
\newblock \bibinfo{title}{Testing the kinetic energy functional: {Kinetic}
  energy density as a density functional}.
\newblock \emph{\bibinfo{journal}{The Journal of Chemical Physics}}
  \textbf{\bibinfo{volume}{118}}, \bibinfo{pages}{8140--8148}
  (\bibinfo{year}{2003}).

\bibitem{BW17}
\bibinfo{author}{Banafsheh, M.} \& \bibinfo{author}{Wesolowski, T.}
\newblock \bibinfo{title}{Nonadditive kinetic potentials from inverted
  kohn--sham problem}.
\newblock \emph{\bibinfo{journal}{International Journal of Quantum Chemistry}}
  (\bibinfo{year}{2017}).

\bibitem{GBV09}
\bibinfo{author}{G{\"{o}}tz, A.~W.}, \bibinfo{author}{Beyhan, S.~M.} \&
  \bibinfo{author}{Visscher, L.}
\newblock \bibinfo{title}{Performance of {Kinetic Energy Functionals} for
  {Interaction Energies} in a {Subsystem Formulation} of {Density Functional
  Theory}}.
\newblock \emph{\bibinfo{journal}{Journal of Chemical Theory and Computation}}
  \textbf{\bibinfo{volume}{5}}, \bibinfo{pages}{3161--3174}
  (\bibinfo{year}{2009}).

\bibitem{WS13}
\bibinfo{author}{Weso{\l}owski, T.~A.} \& \bibinfo{author}{Savin, A.}
\newblock \bibinfo{title}{Non-{Additive Kinetic Energy} and {Potential} in
  {Analytically Solvable Systems} and {Their Approximated Counterparts}}.
\newblock \emph{\bibinfo{journal}{Recent Progress in Orbital-free Density
  Functional Theory}} \textbf{\bibinfo{volume}{6}}, \bibinfo{pages}{275}
  (\bibinfo{year}{2013}).

\bibitem{MOB12}
\bibinfo{author}{Marques, M.~A.}, \bibinfo{author}{Oliveira, M.~J.} \&
  \bibinfo{author}{Burnus, T.}
\newblock \bibinfo{title}{Libxc: A library of exchange and correlation
  functionals for density functional theory}.
\newblock \emph{\bibinfo{journal}{Computer physics communications}}
  \textbf{\bibinfo{volume}{183}}, \bibinfo{pages}{2272--2281}
  (\bibinfo{year}{2012}).

\bibitem{LP87}
\bibinfo{author}{Lee, C.} \& \bibinfo{author}{Parr, R.~G.}
\newblock \bibinfo{title}{{Gauss}ian and other approximations to the
  first-order density matrix of electronic systems, and the derivation of
  various local-density-functional theories}.
\newblock \emph{\bibinfo{journal}{Physical Review A}}
  \textbf{\bibinfo{volume}{35}}, \bibinfo{pages}{2377--2383}
  (\bibinfo{year}{1987}).

\bibitem{OL91}
\bibinfo{author}{Ou-Yang, H.} \& \bibinfo{author}{Levy, M.}
\newblock \bibinfo{title}{Approximate noninteracting kinetic energy functionals
  from a nonuniform scaling requirement}.
\newblock \emph{\bibinfo{journal}{International Journal of Quantum Chemistry}}
  \textbf{\bibinfo{volume}{40}}, \bibinfo{pages}{379--388}
  (\bibinfo{year}{1991}).

\bibitem{PG85}
\bibinfo{author}{Pearson, E.~W.} \& \bibinfo{author}{Gordon, R.~G.}
\newblock \bibinfo{title}{Local asymptotic gradient corrections to the energy
  functional of an electron gas}.
\newblock \emph{\bibinfo{journal}{The Journal of Chemical Physics}}
  \textbf{\bibinfo{volume}{82}}, \bibinfo{pages}{881--889}
  (\bibinfo{year}{1985}).

\bibitem{Per92}
\bibinfo{author}{Perdew, J.~P.}
\newblock \bibinfo{title}{Generalized gradient approximation for the fermion
  kinetic energy as a functional of the density}.
\newblock \emph{\bibinfo{journal}{Physics Letters A}}
  \textbf{\bibinfo{volume}{165}}, \bibinfo{pages}{79--82}
  (\bibinfo{year}{1992}).

\bibitem{VSK98}
\bibinfo{author}{Vitos, L.}, \bibinfo{author}{Skriver, H.~L.} \&
  \bibinfo{author}{Koll{\'{a}}r, J.}
\newblock \bibinfo{title}{Kinetic-energy functionals studied by surface
  calculations}.
\newblock \emph{\bibinfo{journal}{Physical Review B}}
  \textbf{\bibinfo{volume}{57}}, \bibinfo{pages}{12611--12615}
  (\bibinfo{year}{1998}).

\bibitem{VJKS00}
\bibinfo{author}{Vitos, L.}, \bibinfo{author}{Johansson, B.},
  \bibinfo{author}{Koll{\'{a}}r, J.} \& \bibinfo{author}{Skriver, H.~L.}
\newblock \bibinfo{title}{Local kinetic-energy density of the {Airy} gas}.
\newblock \emph{\bibinfo{journal}{Physical Review A}}
  \textbf{\bibinfo{volume}{61}}, \bibinfo{pages}{052511}
  (\bibinfo{year}{2000}).

\bibitem{Ern00}
\bibinfo{author}{Ernzerhof, M.}
\newblock \bibinfo{title}{The role of the kinetic energy density in
  approximations to the exchange energy}.
\newblock \emph{\bibinfo{journal}{Journal of Molecular Structure: {THEOCHEM}}}
  \textbf{\bibinfo{volume}{501–502}}, \bibinfo{pages}{59--64}
  (\bibinfo{year}{2000}).

\end{thebibliography}
